# Unraveling Intrinsic Thermal Conductivity in Layered Conductive MOF Single Crystals

Jin-Kun Guo†, Dong-Yang Wang†, Zhi-Yi Li†, Hao-Yang Zhang, Jia-Xiang Zhang, Ze-Yue Zhang, Lei Sun, Jun-Liang Sun, Jia-Wei Zhou, Chong-An Di, and Jin-Hu Dou*

**ABSTRACT:** Layered conductive metal-organic frameworks (LCMOFs) show great promise in energy and electronics due to their high electrical conductivity and tunable pore structures. They are considered ideal "phonon-glass, electron-crystal" materials. However, their intrinsic thermal transport properties, particularly the thermal conductivity in the single-crystalline state, have never been explored before. The applicability of the Wiedemann-Franz law to such complex porous materials is a key scientific question to describe their thermoelectric relationship. We investigated single crystals of three LCMOFs ($Cu_3HHTP_2$, $Co_9HHTP_4$, $Nd_3HHTP_2$) using the microfabricated suspended device. Results showed ultralow thermal conductivities (0.075-0.194 W $m^{-1}$ $K^{-1}$) along the π-π stacking direction. Crucially, $Nd_3HHTP_2$ exhibited a high electrical conductivity of 398 S $cm^{-1}$, yet its thermal conductivity (0.148 W $m^{-1}$ $K^{-1}$) was comparable to the other two LCMOFs with significantly lower electrical conductivities. Structural characterization revealed that the incommensurate modulation, and in-plane correlated disorder within the $Nd_3HHTP_2$ structure are the potential causes of strong phonon scattering and the observed ultralow thermal conductivity.

## Introduction

Metal-organic frameworks (MOFs), a class tailorable porous materials with periodical structures, are composed of inorganic clusters and organic building blocks.[1-3] MOFs are usually considered as insulating materials, but in recent years, with the deepening understanding of the electronic structure and regulation methods of such materials, conductive metal-organic frameworks have developed rapidly.[4-7] Among them, layered conductive metal-organic frameworks (LCMOFs) have attracted significant attention due to their unique π-conjugated aromatic planar network structures.[8,9] The porous and conductive characteristics make it applicable in fields such as energy storage (such as supercapacitor),[10] gas sensing,[11] electrocatalysis,[12] and so on. These devices are accompanied by significant Joule heating and exothermic reactions during operation, thus thermal conductivity properties of MOFs influence the extent of thermal diffusion.[13] In thermoelectric field, thermal conductivity of LCMOFs directly affects their energy conversion efficiency.[14,15]

Although the thermal conductivity of LCMOFs is an important property, it is noteworthy that current research on heat conduction in such porous materials is relatively scarce, and the mechanisms remain unclear. Recent studies have shown that the porous structure of LCMOFs strongly scatters phonons, resulting in lower thermal conductivity. In 2017, Mircea Dincă et al. reported a thermal conductivity of 0.21 W $m^{-1}$ $K^{-1}$ for pressed pellets of NiHITP (HITP = 2,3,6,7,10,11-hexaiminotriphenylene) at room temperature.[16] In 2020, Daoben Zhu et al. reported a thermal conductivity of 0.2 W $m^{-1}$ $K^{-1}$ for Ni-PTC pellet (PTC = 1,2,3,4,5,6,7,8,9,10,11,12-perthiolated coronene).[17] They also reported a thermal conductivity of 0.17 W $m^{-1}$ $K^{-1}$ for pure Ni-THT film (THT = 1,2,5,6,9,10-triphenylenehexathiol).[18] Overall, films or pellet samples of LCMOFs possess very low thermal conductivity. However, the numerous grain boundaries and defects in these samples cannot reflect the intrinsic thermal transport properties, which hinders the research on the structure-property relationship of its low thermal conductivity.

Herein, we report the first thermal conductivity measurement of LCMOF single crystals. Based on synthesizing high-quality single crystals, the thermal conductivity of the crystal along the π-π stacking direction was tested using a microfabricated suspended device. In order to further investigate the factors affecting heat conduction, we conducted single-crystal X-ray diffraction analysis, and the results showed that the presence of incommensurate modulation and correlated disorder in the crystal played a key role in suppressing thermal conductivity.

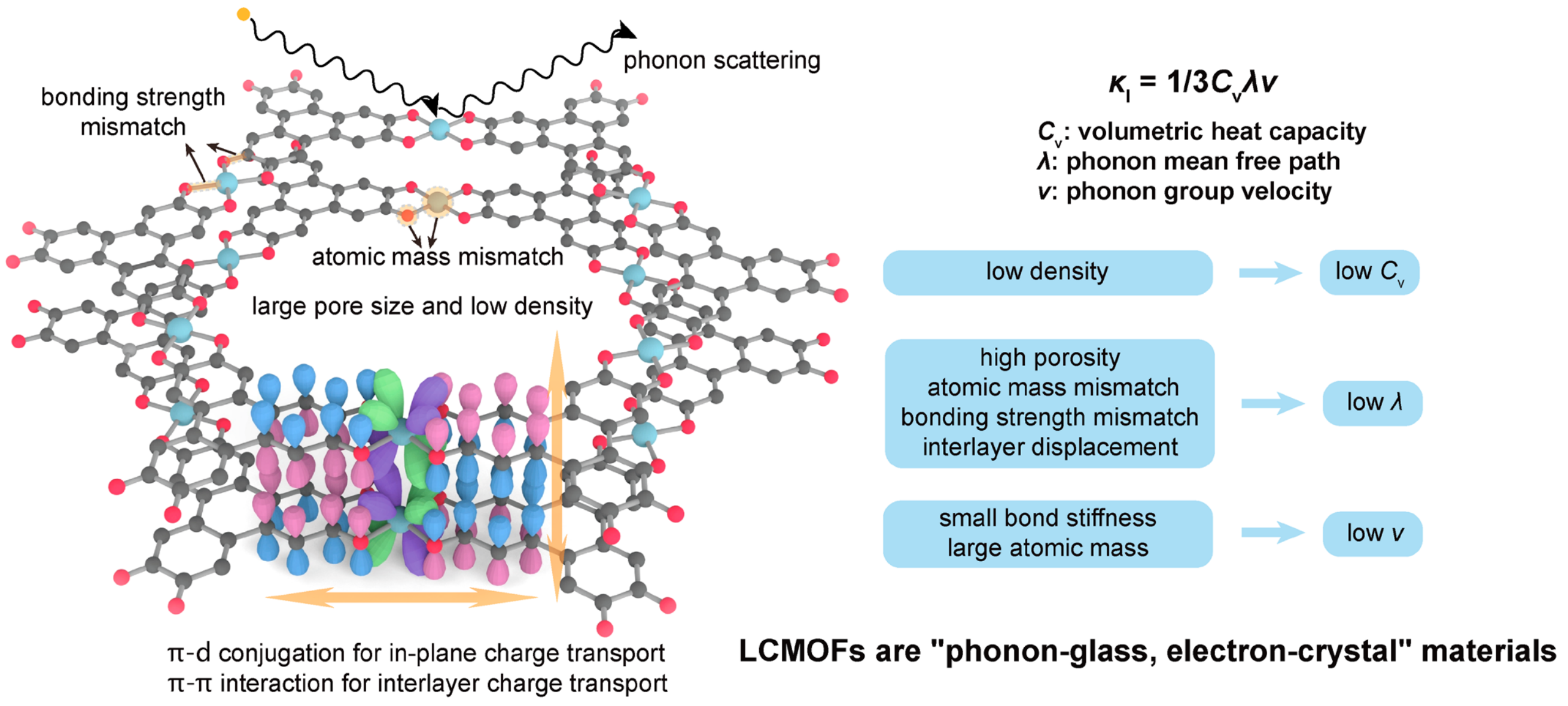

**Figure 1.** Schematic model of layered conductive metal-organic frameworks (LCMOFs). LCMOFs have large pore size and low density; the coexistence of metal and light elements can cause mass mismatch; the coexistence of covalent bonds, coordination bonds, and van der Waals forces can cause bond strength mismatch. These structural factors endow LCMOFs with inherently low thermal conductivity. Meanwhile, LCMOFs can achieve both in-plane and out-of-plane high electrical conductivity, which make them ideal "phonon-glass, electron-crystal" materials.

## Results and Discussion

### Basic theory of solid heat conduction and structural characteristics of LCMOFs

In solids, heat is primarily carried by electrons or phonons. The electronic contribution to thermal conductivity ($\kappa_e$) can be usually approximated by the Wiedemann-Franz law as $\kappa_e = \sigma LT$, where $\sigma$ is the electrical conductivity, $L$ is the Lorenz number, and $T$ is the absolute temperature. This law is derived from the free-electron gas model and is applicable to metals and degenerate semiconductors.[19] However, this law often fails for non-degenerate semiconductors, insulators, as well as low-dimensional materials and strongly correlated systems. For instance, in the organic polymer PEDOT:PSS, the electrical conductivity can be high, but phonons are strongly scattered by structural disorder, resulting in a low thermal conductivity.[20] In strongly correlated electron systems like $CeCu_6$, the effective mass can be thousands of times that of free electrons, rendering them ineffective at carrying heat flow.[21] In these material systems, lattice thermal conductivity ($\kappa_l$) dominates.

The lattice thermal conductivity is given by $\kappa_l = 1/3 C_v \lambda v$, where $C_v$ is the volumetric heat capacity, proportional to the material's density. Lower density results in a smaller volumetric heat capacity. $\lambda$ is the phonon mean free path, representing the average distance that phonons travel between two scattering events; more frequent scattering leads to a shorter mean free path and lower thermal conductivity. Scattering originates from two main aspects: various defects in the material and its intrinsic anharmonicity. Defect-induced scattering includes scattering at interfaces and scattering by defects or impurities (e.g., point defects, dislocations).[22] Lattice anharmonicity refers to the phenomenon where atomic vibrations around equilibrium positions deviate from simple harmonic motion, and the restoring force is no longer strictly proportional to displacement. It is related to the intrinsic structure of the material. Systems with atomic mass mismatch or bonding strength mismatch may exhibit strong anharmonicity.[23] $v$ represents the average phonon group velocity, indicating how fast phonons travel, which is also influenced by intrinsic structural factors of the material. Smaller bond stiffness and larger atomic masses lead to smaller phonon group velocities.

As shown in Figure 1, LCMOFs have low density and high porosity; the coexistence of metal and light elements can cause mass mismatch; the coexistence of covalent bonds, coordination bonds, and van der Waals forces can cause bond strength mismatch, theoretically leading to low lattice thermal conductivity. In-plane, layered conductive MOFs achieve carrier delocalization and efficient charge transport through π-d conjugation, while interlayer charge transport occurs via π-π interactions, resulting in high electrical conductivity. Overall, LCMOFs are ideal "phonon-glass, electron-crystal" materials.

However, the mechanism of intrinsic thermal conductivity of porous materials is currently unclear in the field. LCMOF single crystals provide us with a platform for studying electron and phonon heat conduction in such materials. Therefore, we chose three kinds of LCMOFs—$Cu_3HHTP_2$, $Co_9HHTP_4$, and $Nd_3HHTP_2$—using HHTP (HHTP = 2,3,6,7,10,11-hexahydroxytriphenylene) as the conjugated ligand, and synthesized single crystals of MOFs via a hydrothermal method. The morphology of single crystals was characterized by scanning electron microscope (SEM) and optical microscope (OM) (Figures S1−S6). These images showed that all single crystals possessed rod-like morphologies. Their structures were confirmed by powder X-ray diffraction (PXRD) (Figures S7−S9) while components were confirmed by X-ray photoelectron spectroscopy (XPS) (Figures S16−S18). Fitting the $N_2$ adsorption isotherms of three MOFs to the Brunauer–Emmett–Teller (BET) equation gives apparent surface areas of 239.63 $m^2\ g^{-1}$, 377.48 $m^2\ g^{-1}$, 309.46 $m^2\ g^{-1}$ for $Cu_3HHTP_2$, $Co_9HHTP_4$, and $Nd_3HHTP_2$, respectively (Figures S13−S15).

The structure of $Cu_3HHTP_2$ was obtained by Pawley fitting of experimental PXRD pattern. The structure of $Cu_3HHTP_2$ is shown in Figure 2a, d, g and Table S2. Within the layers, HHTP and $Cu^{2+}$ adopt a square planar coordination, forming a honeycomb tiling structure in plane. The distance between opposite $Cu^{2+}$ ions across the hexagon is 22.0 Å. Adjacent layers are connected by π-π interactions with an interlayer spacing of 3.40 Å. A displacement of 1.12 Å occurs between adjacent layers in the $b$ direction. The theoretical density of $Cu_3HHTP_2$ is 1.0043 g $cm^{-3}$.

The structure of $Co_9HHTP_4$ has been previously reported.[24] The structure of $Co_9HHTP_4$ is shown in Figure 2b, e, h and Table S3. The $Co^{2+}$ coordination number is 6, forming a honeycomb tiling structure in-plane with HHTP. Simultaneously, $Co^{2+}$ can coordinate with water molecules. In the side view, it can be observed that the first and the third layers form continuous structures, while in the second layer, each HHTP coordinates with three $Co^{2+}$ ions to form a cluster. These clusters are arranged parallel to each other without bonding between them. The distance between opposite $Co^{2+}$ ions across the hexagon in the first and third layers is 22.1 Å. The layers are connected by π-π interactions with an average interlayer spacing of 3.33 Å. The theoretical density of $Co_9HHTP_4$ is approximately 1.589 g $cm^{-3}$.

The structure of $Nd_3HHTP_2$ was obtained by single crystal X-ray diffraction (SCXRD). The structure of $Nd_3HHTP_2$ is shown in Figure 2c, f, i. The $Nd^{3+}$ coordination number is 8, forming coordination between adjacent HHTP ligand layers. The distance between opposite $Nd^{3+}$ ions across the hexagon in the same layer is 22.0 Å. The interlayer spacing is only 3.01 Å, the smallest among the three. The theoretical density of $Nd_3HHTP_2$ is approximately 1.733 g $cm^{-3}$.

Furthermore, we observed rod-like single crystals of these three LCMOFs using cryo-electron microscopy (Cryo-EM). In Figure 3d, the spacing between adjacent stripes for $Cu_3HHTP_2$ is 18.4 Å; the first peak in the powder diffraction pattern at $2\theta$ = 4.75° corresponds to the (100) interplanar spacing of 18.6 Å (Figure S7). In Figure 3e, the spacing for $Co_9HHTP_4$ is 19.0 Å; the first peak at $2\theta$ = 4.63° corresponds to the (100) spacing of 19.1 Å (Figure S8). In Figure 3f, the spacing for $Nd_3HHTP_2$ is 19.7 Å; the first peak at $2\theta$ = 4.75° corresponds to the (100) spacing of 19.1 Å (Figure S9). All experimental PXRDs are consistent with the simulation results.

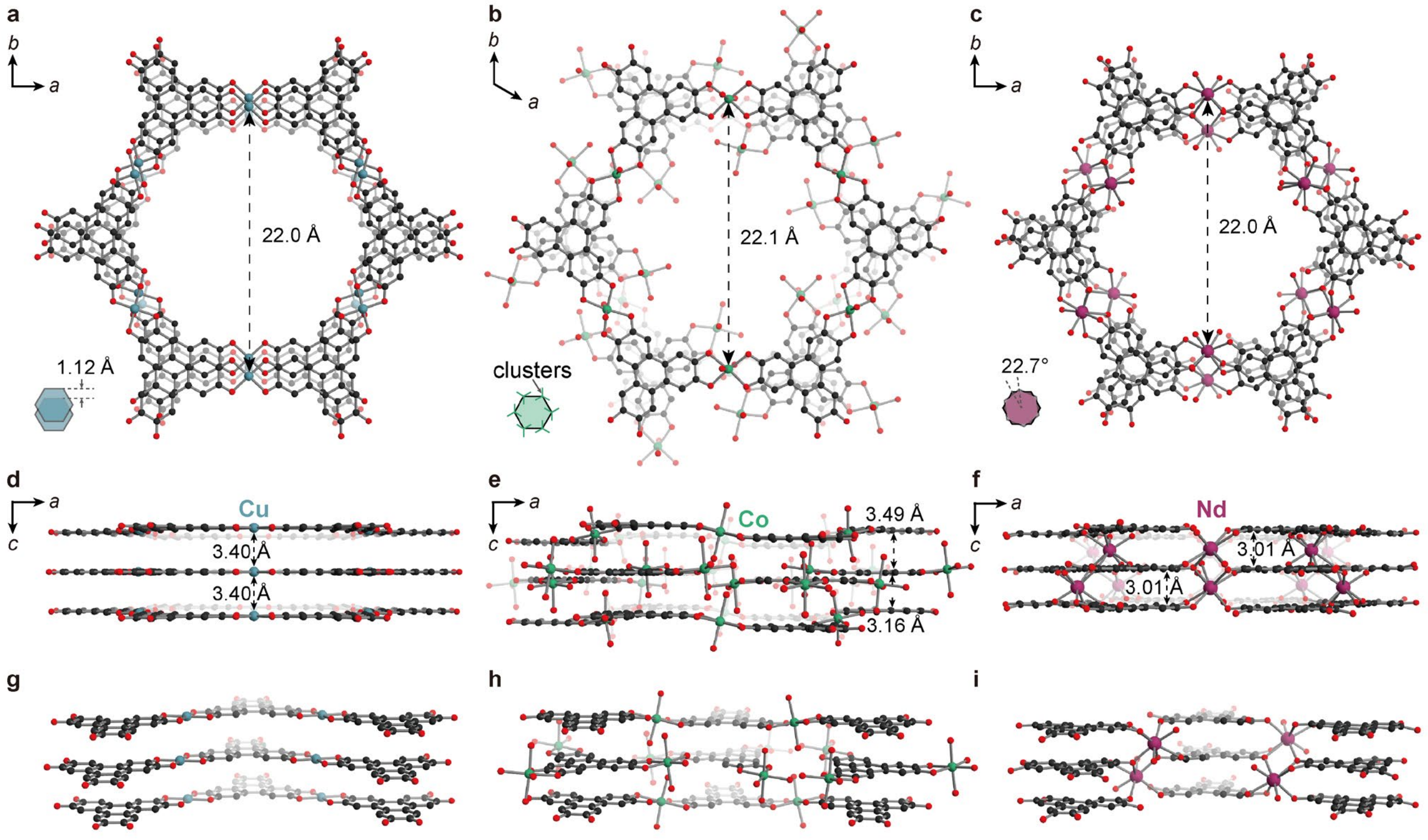

**Figure 2.** (a, b, c) Portions of the crystal structures viewed along the *c* direction for $Cu_3HHTP_2$, $Co_9HHTP_4$, and $Nd_3HHTP_2$ respectively. (d, e, f) Portions of the crystal structures viewed along the *b* direction for $Cu_3HHTP_2$, $Co_9HHTP_4$, and $Nd_3HHTP_2$ respectively. (g, h, i) Side view cross-sections for $Cu_3HHTP_2$, $Co_9HHTP_4$, and $Nd_3HHTP_2$ respectively.

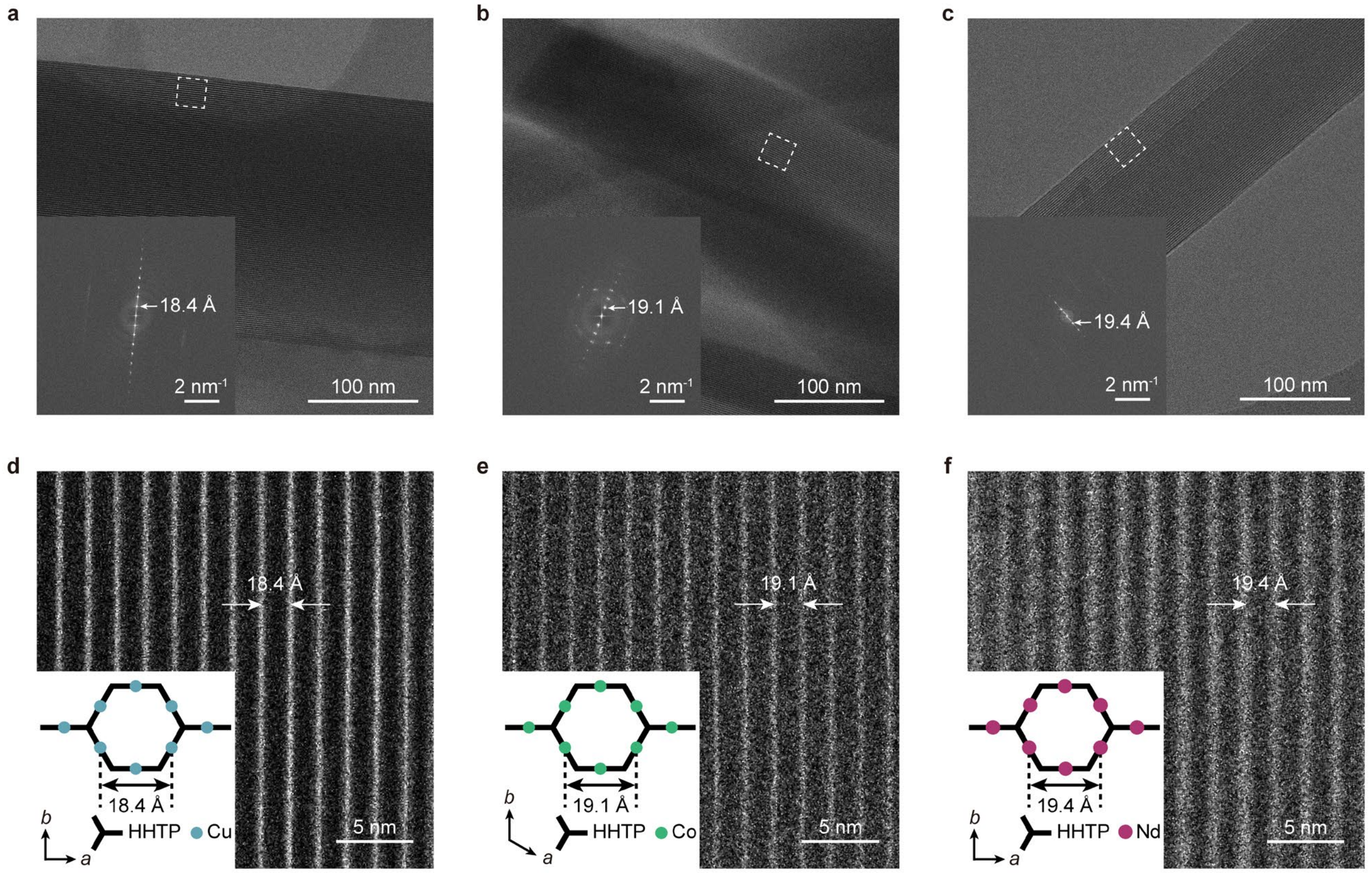

**Figure 3.** (a) Cryo-EM image of $Cu_3HHTP_2$ rod (inset: FFT), (d) is an enlarged view of the white dashed square in (a) (inset: schematic of $Cu_3HHTP_2$ structure to illustrate the source of stripe spacing). (b) Cryo-EM image of a $Co_9HHTP_4$ rods (inset: FFT), (e) is an enlarged view of the white dashed square in (b) (inset: schematic of $Co_9HHTP_4$ structure to illustrate the source of stripe spacing). (c) Cryo-EM image of a $Nd_3HHTP_2$ rod (inset: FFT), (f) is an enlarged view of the white dashed square in (c) (inset: schematic of $Nd_3HHTP_2$ structure to illustrate the source of stripe spacing).

## Electrical and thermal conductivity of LCMOFs

We measured the electrical conductivity of the LCMOF single crystals at room temperature. For the smaller crystals $Cu_3HHTP_2$ and $Co_9HHTP_4$, we fabricated electrodes using electron beam lithography (EBL) and magnetron sputtering, with the device shown in Figure 4c. The process flow diagram is shown in Figure S20. For the larger $Nd_3HHTP_2$ crystal, we transferred single crystals onto a pre-patterned gold electrode array on a $Si/SiO_2$ substrate by single crystal needle and then applied silver paste at the contact points between the crystal and electrodes. The process flow diagram and the device are shown in Figure S21 and Figure S22, respectively. To eliminate contact resistance, we used four-probe method for measurement (Figure S19). The devices and results are shown in Figure 4a, b and Figures S23, S24. The highest electrical conductivities for $Cu_3HHTP_2$ and $Co_9HHTP_4$ are 0.70 S $cm^{-1}$ and 0.69 S $cm^{-1}$, respectively, while that for $Nd_3HHTP_2$ is as high as 398 S $cm^{-1}$. Structurally, $Nd_3HHTP_2$ has the smallest interlayer spacing (3.05 Å), suggesting stronger $\pi$-$\pi$ interactions and higher carrier mobility, leading to higher conductivity. Previous DFT calculations on $Nd_3HHTP_2$ revealed strong band dispersion crossing the Fermi level along the AΓ direction (the $\pi$-$\pi$ stacking direction), while bands in other directions were flat with band gaps. Furthermore, its electrical conductivity along this direction decreased with increasing temperature, indicating metallic electrical transport behavior.[25]

Microfabricated suspended device is often used to measure thermal conductivity across electrode directions. Such devices have been widely applicable to rod-shaped crystal and thin film samples.[27-29] Herein, we measured intrinsic thermal conductivity of LCMOFs single crystals along the $\pi$-$\pi$ stacking direction by this method at room temperature.

We transferred the three MOF single crystals onto microfabricated suspended device using a Focused Ion Beam (FIB) mechanical arm and fixed the contacts by depositing Pt. The diagram of device fabrication process and SEM image of the device are shown in Figures 4d, S25−S28. The results of thermal conductivity, shown in Figure 4e, indicate that the thermal conductivity decreases in the order $Co_9HHTP_4$, $Nd_3HHTP_2$, $Cu_3HHTP_2$, with values of 0.194 ± 0.004 W $m^{-1}$ $K^{-1}$, 0.148 ± 0.002 W $m^{-1}$ $K^{-1}$, and 0.075 ± 0.004 W $m^{-1}$ $K^{-1}$ $K^{-1}$, respectively (Figure 4e). Notably, although the single-crystal electrical conductivity of $Nd_3HHTP_2$ is three orders of magnitude higher than the others, its thermal conductivity is comparable. As mentioned, $Nd_3HHTP_2$ exhibits metallic electrical transport along the $\pi$-$\pi$ stacking direction with high conductivity (398 S $cm^{-1}$). Calculating $\kappa_e$ using the Wiedemann-Franz law gives a theoretical value range from 0.226 W $m^{-1}$ $K^{-1}$ to 0.289 W $m^{-1}$ $K^{-1}$ (corresponding to electrical conductivities from 311 S $cm^{-1}$ to 398 S $cm^{-1}$) at room temperature, which is already higher than the measured total thermal conductivity (Figure 4e). This demonstrates that $Nd_3HHTP_2$ deviates from the Wiedemann-Franz law at room temperature. The significant difference in electrical conductivity but small difference in thermal conductivity among the LCMOFs proves that lattice thermal conductivity dominates in such material system. Below, we analyze potential influencing factors from a structural perspective.

## Structural origins of low thermal conductivity in $Nd_3HHTP_2$

We first analyzed $Nd_3HHTP_2$ single crystals using single crystal X-ray diffraction. Figure 5a shows room-temperature single-crystal X-ray diffraction pattern for $Nd_3HHTP_2$ which is vertical to $a^*$ direction. In (*hkl*) lattice plane, when $l$ is even, satellite diffraction spots appear besides each row of main diffraction points (marked by white dashed box), a typical feature of a modulated structure, also observed in previous studies.[25,26,30] A modulated structure refers to a structure where atoms periodically deviate from their average positions. When the period of modulation is commensurate with the lattice period, it's a commensurate modulation; otherwise, it's incommensurate. $Nd_3HHTP_2$ exhibits an incommensurate modulation along the $c$ direction (the $\pi$-$\pi$ stacking direction), with a modulation vector of approximately $0.393c^*$. Figure 5d shows fluctuations in the interlayer spacing of $Nd_3HHTP_2$ along the $c$ direction. Furthermore, structural modulation likely indicates the presence of strong electron-phonon coupling in materials. The strong electron-phonon coupling might cause Kohn anomaly, where a phonon mode with a specific wave vector $q = 2k_F$ ($k_F$ is the Fermi wave vector) becomes unstable and strongly scatters electrons near the Fermi surface. The vibrational frequencies of these phonons decrease and may even approach zero, a phenomenon known as phonon softening. Once the lattice becomes unstable at $q = 2k_F$, the system lowers its total energy by opening an electronic energy gap and developing a period-doubled structural distortion, known as a Peierls distortion.[31-33] Figure 5b illustrates the formation of Peierls distortion in a one-dimensional metal chain. Under normal conditions, atoms are uniformly distributed with lattice constant $a$, and the charge density $\rho$(x) along the chain is uniform. When a Peierls distortion occurs, neighboring atoms move closer, doubling the lattice period to $2a$, and $\rho$(x) exhibits fluctuations.

Similarly, the incommensurate modulation observed in $Nd_3HHTP_2$ can also originate from phonon softening. However, Unlike the Peierls case where the soft phonon emerges exactly at $q = 2k_F$, the total energy of $Nd_3HHTP_2$ is minimized at a wave vector that does not lock to an integer multiple of the underlying lattice periodicity. This leads to an irrational modulation period and, consequently, an incommensurate modulation. In summary, the presence of an incommensurate modulation embedded in the $Nd_3HHTP_2$ framework is a major reason for its deviation from the Wiedemann-Franz law and intrinsic low thermal conductivity.

Selected area electron diffraction (SAED) was also made for $Nd_3HHTP_2$ single crystals and Figure 5c shows the image. When $l$ is odd, the diffraction spots exhibit elongation (marked by white arrows), indicating diffuse scattering caused by correlated disorder within the *ab*-plane. Correlated disorder refers to disorder in some dimensions while maintaining order in others.[34] It's shown in Figure 5d that between adjacent layers, $Nd^{3+}$ ions have two different occupation sites, forming a complementary pattern. The blue and purple polyhedra represent these two different occupation sites and only one of them exists in this column. Each type of site forms a zigzag chain between the layers. Within the *ab*-plane, due to the randomness of the two site occupations, the distribution of $Nd^{3+}$ exhibits correlated disorder. Correlated disorder can be considered an additional scattering source, contributing to phonon scattering and the observed low thermal conductivity.

We compiled the room-temperature electrical and thermal conductivities of some representative materials, as shown in Figure 4f. Among all materials with electrical conductivity above 100 S $cm^{-1}$, $Nd_3HHTP_2$ possesses the lowest thermal conductivity. And $Nd_3HHTP_2$ tests the thermal conductivity of single crystals, while other materials are pellet or thin film samples. This value was measured on single crystals, minimizing interference from various defects and reflecting more intrinsic thermal transport properties. The bottom right corner of the figure highlights materials simultaneously possessing high electrical conductivity and low thermal conductivity, which have wide applications in thermoelectric conversion, energy storage, etc., such as the inorganic material $Sb_2Te_3$ and the organic semiconductor PEDOT:PSS. $Nd_3HHTP_2$'s performance is comparable to these, suggesting its potential as a new candidate in the field of thermoelectric materials.[35,36]

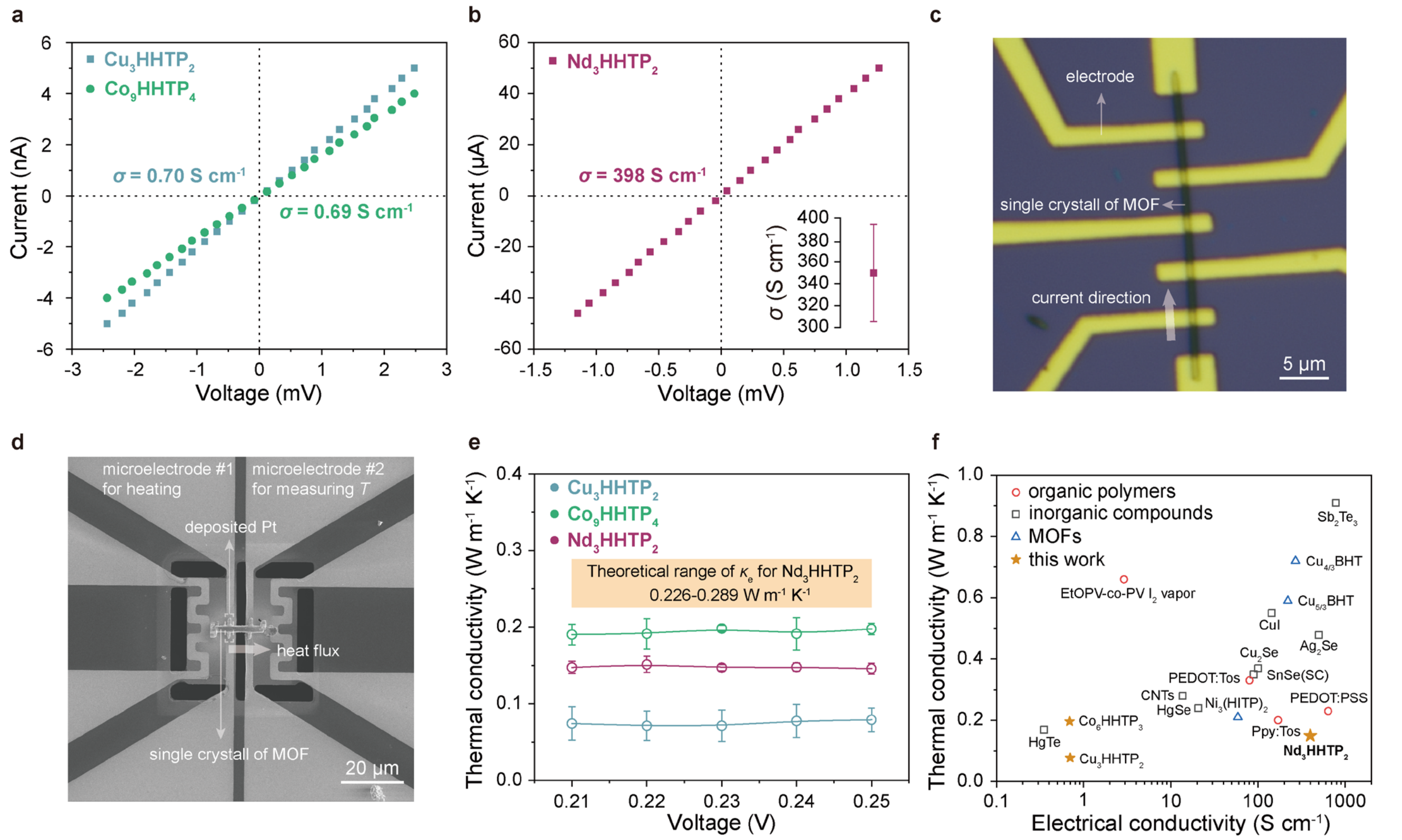

**Figure 4.** (a) *I−V* curves of $Cu_3HHTP_2$ (blue) and $Co_9HHTP_4$ (green). (b) *I−V* curves of $Nd_3HHTP_2$. (Inset: error bar of three samples) (c) OM image of EBL device for electrical conductivity measurement. (d) SEM image of microfabricated suspended device for thermal conductivity measurement. (e) Thermal conductivity of $Cu_3HHTP_2$ (blue), $Co_9HHTP_4$ (green) and $Nd_3HHTP_2$ (purple) along *c* direction. (f) Comparison of the thermal conductivities with electrical conductivities for organic polymers, inorganic compounds and MOFs (including three MOFs in this work). Only SnSe and three MOFs in this work tested the thermal conductivity of single crystal samples, while the rest were pellet or thin film samples.

## Conclusions

In summary, we have conducted the first study on the intrinsic thermal conductivity of LCMOF single crystals. By synthesizing high-quality single crystals of $Cu_3HHTP_2$, $Co_9HHTP_4$, and $Nd_3HHTP_2$, we directly measured their thermal conductivity along the π-π stacking direction using a microfabricated suspended device. All three LCMOFs exhibit ultralow thermal conductivity (0.075–0.194 W $m^{-1}$ $K^{-1}$), consistent with their structural characteristics of low density, high porosity, and significant atomic mass and bond-strength mismatch, which strongly scatter phonons.

Notably, $Nd_3HHTP_2$ demonstrates a remarkable combination of metallic electrical conductivity (~398 S $cm^{-1}$) and an exceptionally low thermal conductivity (~0.148 W $m^{-1}$ $K^{-1}$), which significantly deviates from the Wiedemann-Franz law. Through single-crystal X-ray diffraction and electron diffraction analysis, we attribute this unique behavior to the presence of an incommensurate modulation and correlated disorder within its crystal structure. The incommensurate modulation, indicative of strong electron-phonon coupling and potential phonon softening, alongside the *ab*-plane correlated disorder of $Nd^{3+}$, act as intrinsic scattering centers that drastically suppress phonon transport while preserving high electronic mobility.

These findings establish LCMOF single crystals, particularly $Nd_3HHTP_2$, as a novel class of "phonon-glass, electron-crystal" materials. Their ability to intrinsically decouple electron and phonon transport, as revealed in single-crystalline form, provides fundamental insights into thermal management in porous conductive materials and highlights their significant potential for applications in thermoelectrics and other technologies where high electrical conductivity must coexist with low thermal conductivity.

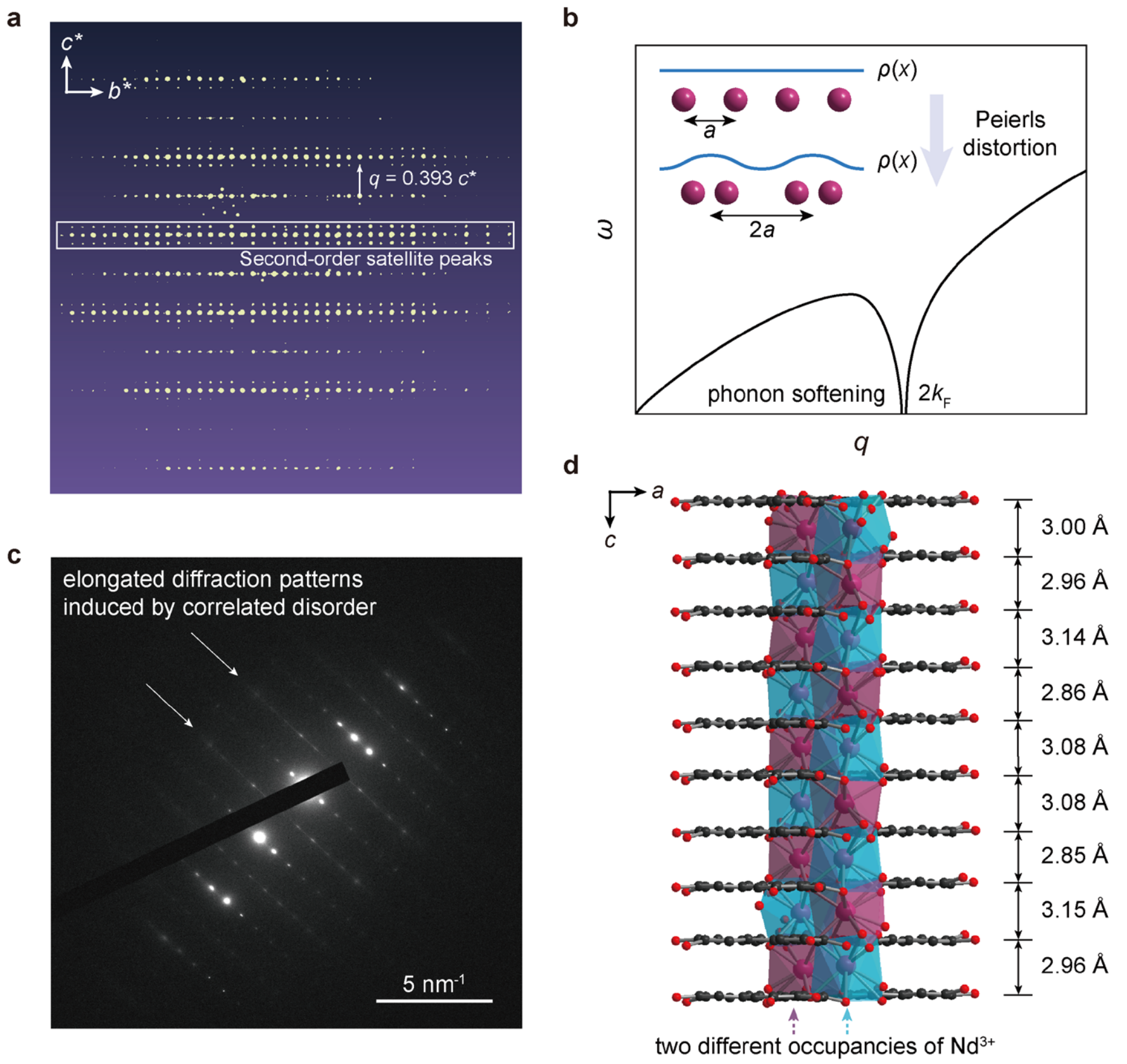

**Figure 5.** (a) Room-temperature single-crystal X-ray diffraction pattern for $Nd_3HHTP_2$ which is vertical to *a** direction, satellite diffraction spots are marked by white dashed box, the modulation vector $q = 0.393\ c^*$. (b) Schematic diagram of Peierls distortion in one-dimensional atomic chain and the process of phonon softening at $q = 2k_F$, $\rho(x)$ refers to the distribution of charge density. (c) SAED image of $Nd_3HHTP_2$ (the elongated diffraction point indicated by the white arrow is caused by correlated disorder). (d) Schematic diagram of $Nd_3HHTP_2$ structure. Along *c* direction (the π-π stacking direction), there are fluctuations in the interlayer spacing, exhibiting incommensurate modulation along this direction. The blue and purple polyhedra represent two different occupancies of $Nd^{3+}$. In this column, $Nd^{3+}$ will randomly adopt only one of the occupancy modes. The randomness of occupation causes the correlated disorder.

## Associated Content

### Supporting Information

The Supporting Information includes Materials and Methods; SEM, OM images of LCMOFs; experimental and simulated PXRD patterns, SAED patterns, nitrogen adsorption isotherms, XPS spectra, crystal data of LCMOFs; schematic diagram of electrical and thermal conductivity measurement devices and results of LCMOFs.

## Author Information

### Corresponding Authors

**Jin-Hu Dou** − National Key Laboratory of Advanced Micro and Nano Manufacture Technology; Key Laboratory of Polymer Chemistry and Physics of Ministry of Education, School of Materials Science and Engineering, Peking University, Beijing, 100871, China. Email: doujinhu@pku.edu.cn

### Authors

**Jin-Kun Guo** − National Key Laboratory of Advanced Micro and Nano Manufacture Technology; Key Laboratory of Polymer Chemistry and Physics of Ministry of Education, School of Materials Science and Engineering, Peking University, Beijing, 100871, China.

**Dong-Yang Wang** − Beijing National Laboratory for Molecular Sciences, CAS Key Laboratory of Organic Solids, Institute of Chemistry, Chinese Academy of Sciences, Beijing 100190, China.

**Zhi-Yi Li** − Beijing National Laboratory for Molecular Sciences, CAS Key Laboratory of Organic Solids, Institute of Chemistry, Chinese Academy of Sciences, Beijing 100190, China.

**Hao-Yang Zhang** − National Key Laboratory of Advanced Micro and Nano Manufacture Technology; Key Laboratory of Polymer Chemistry and Physics of Ministry of Education, School of Materials Science and Engineering, Peking University, Beijing, 100871, China.

**Jia-Xiang Zhang** − National Key Laboratory of Advanced Micro and Nano Manufacture Technology; Key Laboratory of Polymer Chemistry and Physics of Ministry of Education, School of Materials Science and Engineering, Peking University, Beijing, 100871, China.

**Ze-Yue Zhang** − College of Chemistry and Molecular Engineering, Beijing National Laboratory for Molecular Sciences, Peking University, Beijing 100871, China

**Lei Sun** − Department of Chemistry, School of Science and Research Center for Industries of the Future, Westlake University, Hangzhou, Zhejiang Province 310030, China

**Jun-Liang Sun** − College of Chemistry and Molecular Engineering, Beijing National Laboratory for Molecular Sciences, Peking University, Beijing 100871, China

**Jia-Wei Zhou** − Department of Mechanical Engineering, The University of Hong Kong, Pokfulam, Hong Kong SAR, China

**Chong-An Di** − Beijing National Laboratory for Molecular Sciences, CAS Key Laboratory of Organic Solids, Institute of Chemistry, Chinese Academy of Sciences, Beijing 100190, China.

Author Contributions

J.-K.G. and J.-H.D. conceived the idea and designed experiments. J.-K.G. synthesized the single crystals, performed basic characterization including OM, SEM, TEM, PXRD, BET, XPS and so on. J.-K.G. and H.-Y.Z. tested the electrical conductivity of single crystals. J.-K.G., D.-Y.W. and Z.-Y.L. performed FIB sample preparation and thermal conductivity measurement. J.-X.Z. performed single crystal X-ray diffraction test. Z.-Y.Z. assisted in the analysis and refinement of $Nd_3HHTP_2$ incommensurate modulation structure. All authors interpreted the results and wrote the manuscript. J.-H.D. supervised the whole project. These authors contributed equally: J.-K.G., D.-Y.W. and Z.-Y.L.

## Funding Sources

This work was financially supported by the National Key R&D Program of China (Grant No. 2023YFE0206400), National Natural Science Foundation of China (Grant No. 22171185, 22575005), and the Clinical Medicine Plus X-Young Scholars Project at Peking University backed by the Fundamental Research Funds for the Central Universities.

## Acknowledgements

The authors acknowledge Molecular Materials and Nanofabrication Laboratory (MMNL) of Peking University for the use of instruments. Cryo-EM data were collected on the Electron Microscopy Laboratory of Peking University with the assistance of Xue-Mei Li.

Insert Table of Contents artwork here

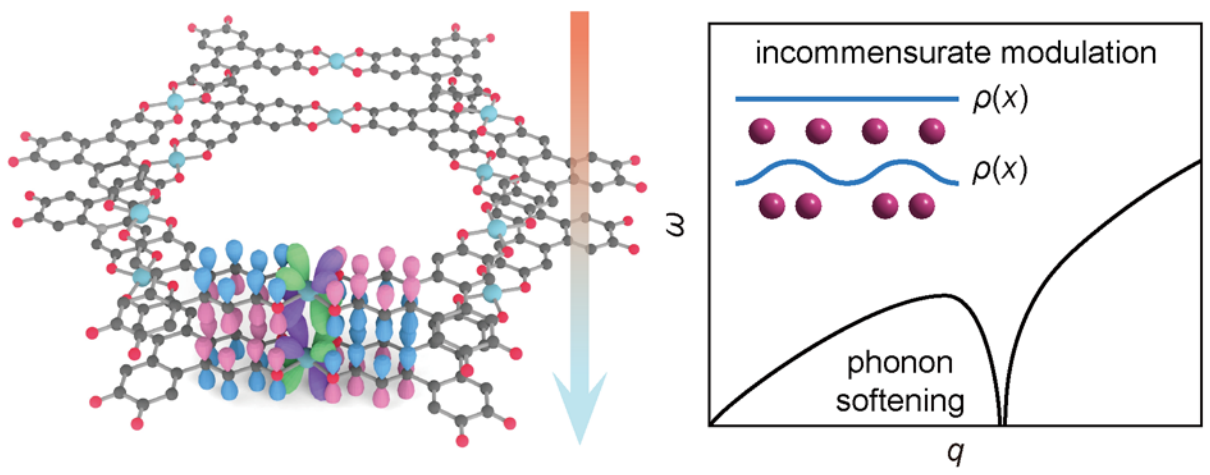

Intrinsic low cross-plane thermal conductivity in layered conductive MOF single crystals.

# Supporting Information for

## Unraveling Intrinsic Thermal Conductivity in Layered Conductive MOF Single Crystals

Jin-Kun Guo[1, †], Dong-Yang Wang[2, †], Zhi-Yi Li[2, †], Hao-Yang Zhang[1], Jia-Xiang Zhang[1], Ze-Yue Zhang[3], Lei Sun[4], Jun-Liang Sun[3], Jia-Wei Zhou[5], Chong-An Di[2], and Jin-Hu Dou[1, *]

[1] National Key Laboratory of Advanced Micro and Nano Manufacture Technology; Key Laboratory of Polymer Chemistry and Physics of Ministry of Education, School of Materials Science and Engineering, Peking University, Beijing, 100871, China.

[2] Beijing National Laboratory for Molecular Sciences, CAS Key Laboratory of Organic Solids, Institute of Chemistry, Chinese Academy of Sciences, Beijing 100190, China

[3] College of Chemistry and Molecular Engineering, Beijing National Laboratory for Molecular Sciences, Peking University, Beijing 100871, China

[4] Department of Chemistry, School of Science and Research Center for Industries of the Future, Westlake University, Hangzhou, Zhejiang Province 310030, China

[5] Department of Mechanical Engineering, The University of Hong Kong, Pokfulam, Hong Kong SAR, China

[†]These authors contributed equally: Jin-Kun Guo, Dong-Yang Wang and Zhi-Yi Li.

[*]Author to whom correspondence should be addressed.

E-mail: doujinhu@pku.edu.cn

**The PDF file includes:**

Materials and Methods
Figs. S1 to S28
Tables S1 to S3
References

**Materials and Methods**

Materials

$Cu(CH_3COO)_2$•$H_2O$ (Sigma-Aldrich, 99%), $Co(CH_3COO)_2$•$4H_2O$ (Sigma-Aldrich, 99%), $Nd(NO_3)_3$•$6H_2O$ (Aladdin, 99.9%), 2,3,6,7,10,11-Hexahydroxytriphenylene (HHTP; Aladdin, 95%), N,N-dimethylformamide (DMF; Sigma-Aldrich, 99%) and N,N-dimethylacetamide (DMA; Sigma-Aldrich, 99%) were used as received. Reagent-grade water was deoxygenated by bubbling with $N_2$ for at least 24 h.

Methods

Growth of $Cu_3HHTP_2$

11.6 mg HHTP was carefully weighed to a 10 mL glass vial, to which 1.5 mL $H_2O$ and 0.2 mL DMF was added. 11.8 mg $Cu(CH_3COO)_2$•$H_2O$ was dissolved in 0.5 mL $H_2O$. Both solutions were ultrasonicated for 2 min and heated at 85 °C for 10 min. Then, they were mixed and heated for 24 h. The product is a black flocculent precipitate.

Growth of $Co_9HHTP_4$

11.6 mg HHTP was carefully weighed to a 10 mL glass vial, to which 1.5 mL $H_2O$ and 0.2 mL DMF was added. 17.8 mg $Co(CH_3COO)_2$•$4H_2O$ was dissolved in 0.5 mL $H_2O$. Both solutions were ultrasonicated for 2 min and heated at 85 °C for 10 min. Then, they were mixed and heated for 24 h. The product is a black flocculent precipitate.

Growth of $Nd_3HHTP_2$

In an $N_2$-filled glove box, 27 mg HHTP was dissolved in 0.7 mL DMA and 1095 mg $Nd(NO_3)_3$•$6H_2O$ was dissolved in 6 mL deoxygenated water by heating to 90°C for 10 min. Both solutions were mixed and transferred into a pressure tube. Frosted glass sheet was put into the tube to improve the crystallinity and yield of the product. The pressure tube was then fitted with a polytetrafluoroethylene screw plug with a back-seal silicone O-ring, taped with electrical tape, and heated at 135 °C for 72 h. The product is a black flocculent precipitate.

Powder X-ray diffraction (PXRD)

Laboratory powder X-ray diffraction patterns of all samples were collected on Rigaku MiniFlex 600 equipped with Ni-filtered CuKα radiation ($K\alpha_1$ = 1.5406 Å, $K\alpha_2$ = 1.5444 Å, $K\alpha_1/K\alpha_2$ = 0.5). A scan rate of 10°/min and a step size of 0.01° was adopted. Samples were prepared by placing a thin layer of the appropriate material on a quartz plate.

Optical microscope (OM)

OM images were obtained by Nikon LV100ND optical microscope.

$N_2$ Adsorption-desorption analysis

The nitrogen adsorption measurements were performed on a Micromeritics ASAP 2020 Surface Area and Porosity Analyzer. An oven-dried sample tube equipped with a TranSeal™ (Micromeritics) was evacuated and tared. The sample was transferred to the sample tube, which was then capped with a TranSeal™. The sample was heated to 80 °C, under a dynamic vacuum of 4 mTorr until the outgas rate was less than 2 mTorr/minute. The evacuated sample tube was weighed again and the sample mass was determined by subtracting the mass of the previously tared tube. The $N_2$ isotherm was measured using a liquid nitrogen bath (77 K). Ultrahigh purity grade (99.999% purity) $N_2$, oil-free valves and gas regulators were used for all the free space correction and measurement. Fits to the Brunauer-Emmett-Teller (BET) [1] equation satisfied the published consistency criteria [2].

X-ray photoelectron spectroscopy (XPS)

XPS was conducted on Thermo Fisher Nexsa X-ray photoelectron spectroscopy.

Cryo-high resolution transmission electron microscopy (Cryo-HRTEM)

Cryo-HRTEM images were obtained with FEI Titan Krios TEM (Gatan K3 summit camera), operated at an accelerating voltage of 300 kV with GIF Quantum energy filter (Gatan). Samples were drop-cast onto Cu TEM grids from powder dispersed in methanol. All image acquisition and analysis of the raw HRTEM data (.dm4) were done using *Digital Micrograph 4.0*.

Single crystal X-ray diffraction test and analysis (SCXRD)

High-quality crystals of $Nd_3HHTP_2$ with suitable sizes were selected for SCXRD analysis. The single-crystal diffraction data collections were collected by using FR-X Mo Kα radiation (λ = 0.7107 Å) at 300 K on Rigaku XtaLAB Synergy Custom diffractometers install with a hybrid photon counting X-ray detector with *CrysAlisPro* [3]. The experimental data of $Nd_3HHTP_2$ are processed using *CrystAlisPro* [3], including peak hunting, intensity integration and indexing of main and satellite reflections. The structural solution and refinement process is then conducted using *JANA2020* [4]. The structures are solved via *Superflip* [5], positions of Nd and part of the atoms in the ligands is determined and the initial density distribution on the extra dimension caused by modulation is generated simultaneously. The structure model is then completed in the refinement process. According to the cumulative distribution function of normalized structure factors (low $|E^2 - 1|$ value) and the needle-shaped nature of the crystal, the merohedral twin domains are added. The refinement is conducted firstly against the main reflections, and then the satellite reflections are included. Disordered water molecules are distributed in the pores and they are represented by multiple oxygen positions with low occupancies in the final structures. Two modulation waves upon position perturbance are defined for each atom.

Single crystal device fabrication for room-temperature electrical conductivity measurement

Samples of $Cu_3HHTP_2$ and $Co_9HHTP_4$ were suspended in methanol by ultrasonication and drop-cast on a pre-patterned $Si/SiO_2$ (300 nm $SiO_2$) substrate followed by spin-coating to evaporate the solvent and disperse the single crystals. Optical alignment and registration were conducted for further lithography. 2 layers of MMA and 1 layer of PMMA was coated on the substrate, where the layers were baked at 150 °C and 170 °C for 2 min, respectively. Patterning was performed by electron beam lithography at 30 kV. 5 nm Ni and 200 nm Au were deposited on substrates by magnetron sputtering. Subsequently, the sample was soaked in acetone for a 24 h-lift-off process.

Samples of $Nd_3HHTP_2$ were suspended in water and drop-cast on a glass slide. Use a single crystal needle to pick up single crystals from water onto a $Si/SiO_2$ (300 nm $SiO_2$) substrate with pre-evaporated gold electrodes. Stick silver paste on the contact area between the crystal and the electrodes to reduce contact resistance.

All 4-probe room-temperature electrical conductivity measurements of single crystal devices were performed using B1500A Semiconductor Device Parameter Analyzer in the dark. *I-V* curves were collected by scanning current with at least 101 steps and measuring voltage at each step.

Single crystal device fabrication for room-temperature thermal conductivity measurement

Thermal conductivity was measured using a microfabricated suspended device based on Fourier's law. Drop the methanol solution of MOFs onto the surface of the 1 cm × 1 cm silicon wafer and spin-dry the solution using a spin coater to distribute the rod-like crystals on the surface of the silicon wafer. Use the mechanical arm of ThermoFisher FIB Helios G4 UX to lift the crystal from the silicon wafer and place it between two suspended electrodes.

The measurement setup consisted of two lock-in amplifiers, two adjustable resistors, two differential amplifiers, and a measurement terminal. The signal output and detection were carried out through separate electrical channels. All measurements were performed at room temperature under high vacuum (~$10^{-4}$ Pa). During measurement, one microelectrode was used as the heating terminal, while the other served as the sensing terminal to record the temperature rise induced by heat flow through the suspended sample. Lock-in amplifiers provided precise heating control and temperature detection, allowing accurate determination of the heat flux and temperature difference across the sample.The thermal conductivity ($\kappa$) was calculated according to Fourier's law:

$$\kappa = \frac{QL}{A\Delta T} \tag{1}$$

where $Q$, $L$, $A$ and $\Delta T$ represent the heat transfer power, heat transfer length, cross-sectional area, and temperature difference between the two electrodes, respectively. The quantities $Q$ and $\Delta T$ were obtained using the following relations:

$$Q = \frac{\Delta T_2}{R_S} = \frac{\Delta T_2 \kappa_{Au} A_{Au}}{L_{Au}} \tag{2}$$

$$\Delta \mathrm{T} = \Delta T_1 - \Delta T_2 \tag{3}$$

$$\Delta T_1 = \frac{\frac{R_1}{R_{01}} - 1}{\alpha} = \frac{\frac{U_{A1}}{U_{B1}} - 1}{\alpha} \tag{4}$$

$$\Delta T_2 = \frac{\frac{R_2}{R_{02}} - 1}{\alpha} = \frac{\frac{U_{A2}}{U_{B2}} - 1}{\alpha} \tag{5}$$

Here, $R_s$, $\kappa_{Au}$, $A_{Au}$ and $L_{Au}$ are the thermal resistance, thermal conductivity, cross-sectional area and length of the gold electrode, respectively. $R_{0i}$ and $R_i$ denote the initial and heated resistances of the electrode, while $U_{Ai}$, and $U_{Bi}$ represent the voltages across the corresponding electrodes and resistors. $\alpha$ is the temperature coefficient of resistance of the electrode material.

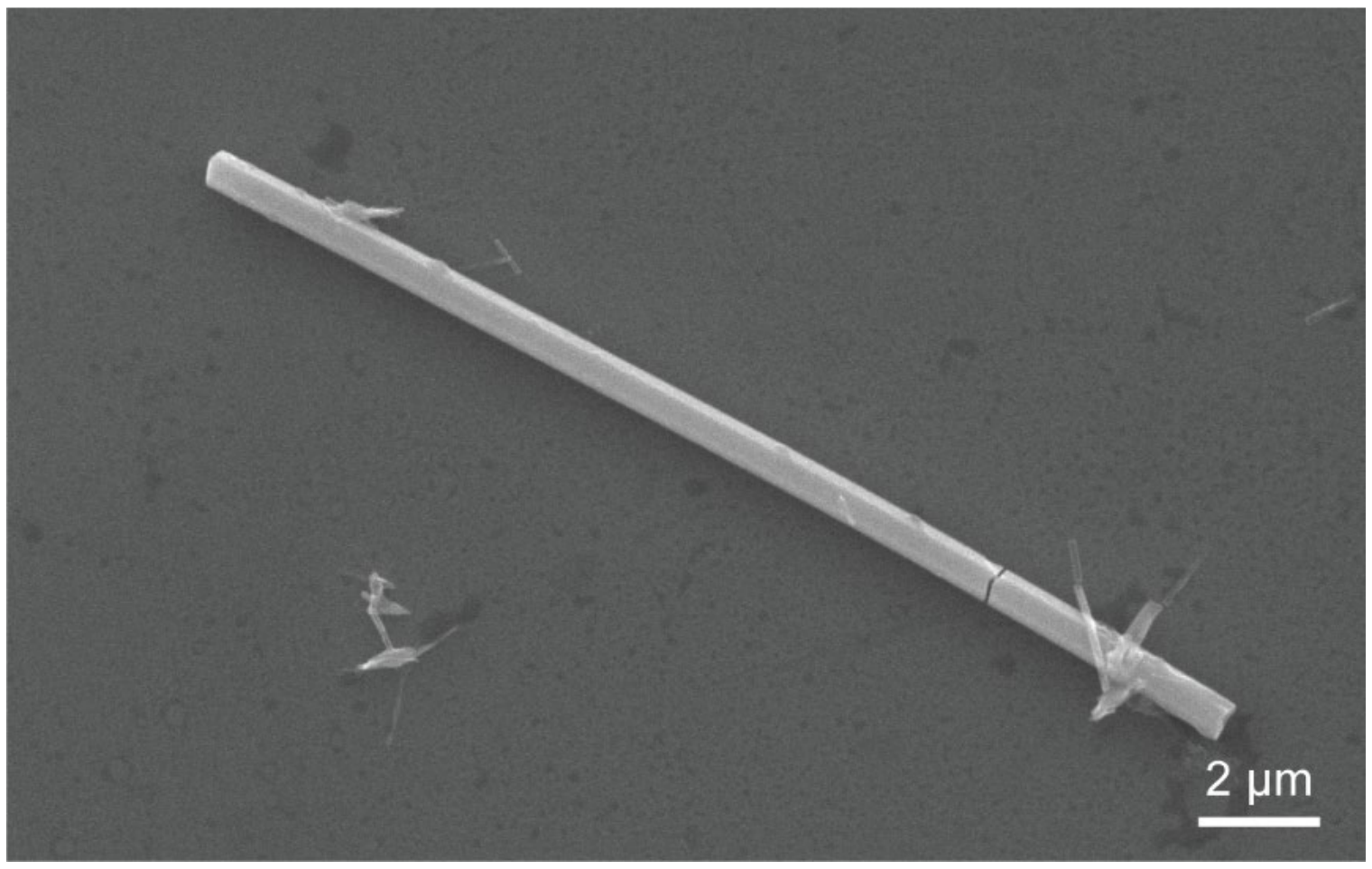

**Fig. S1.**
SEM image of $Cu_3HHTP_2$.

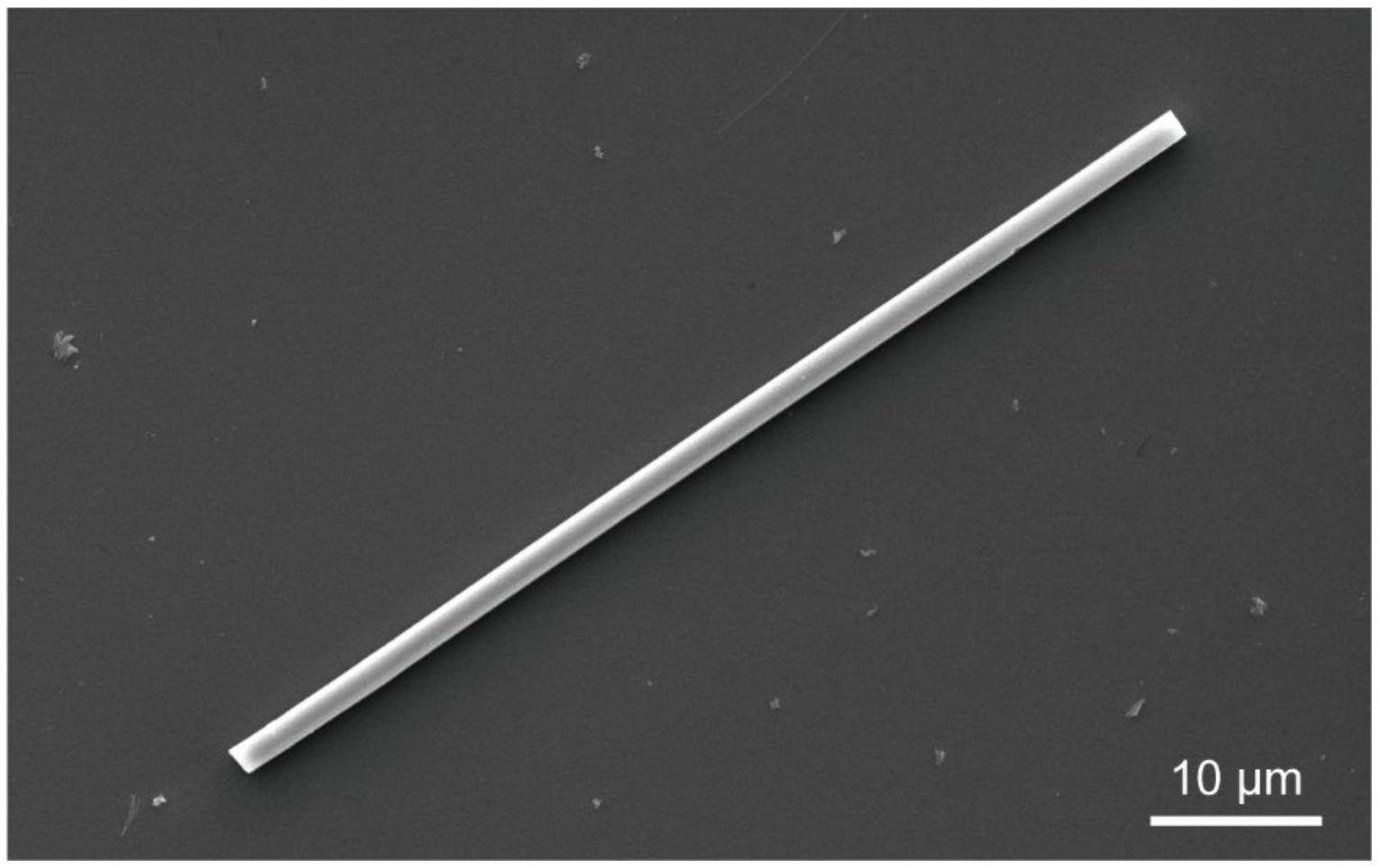

**Fig. S2.**
SEM image of $Co_9HHTP_4$.

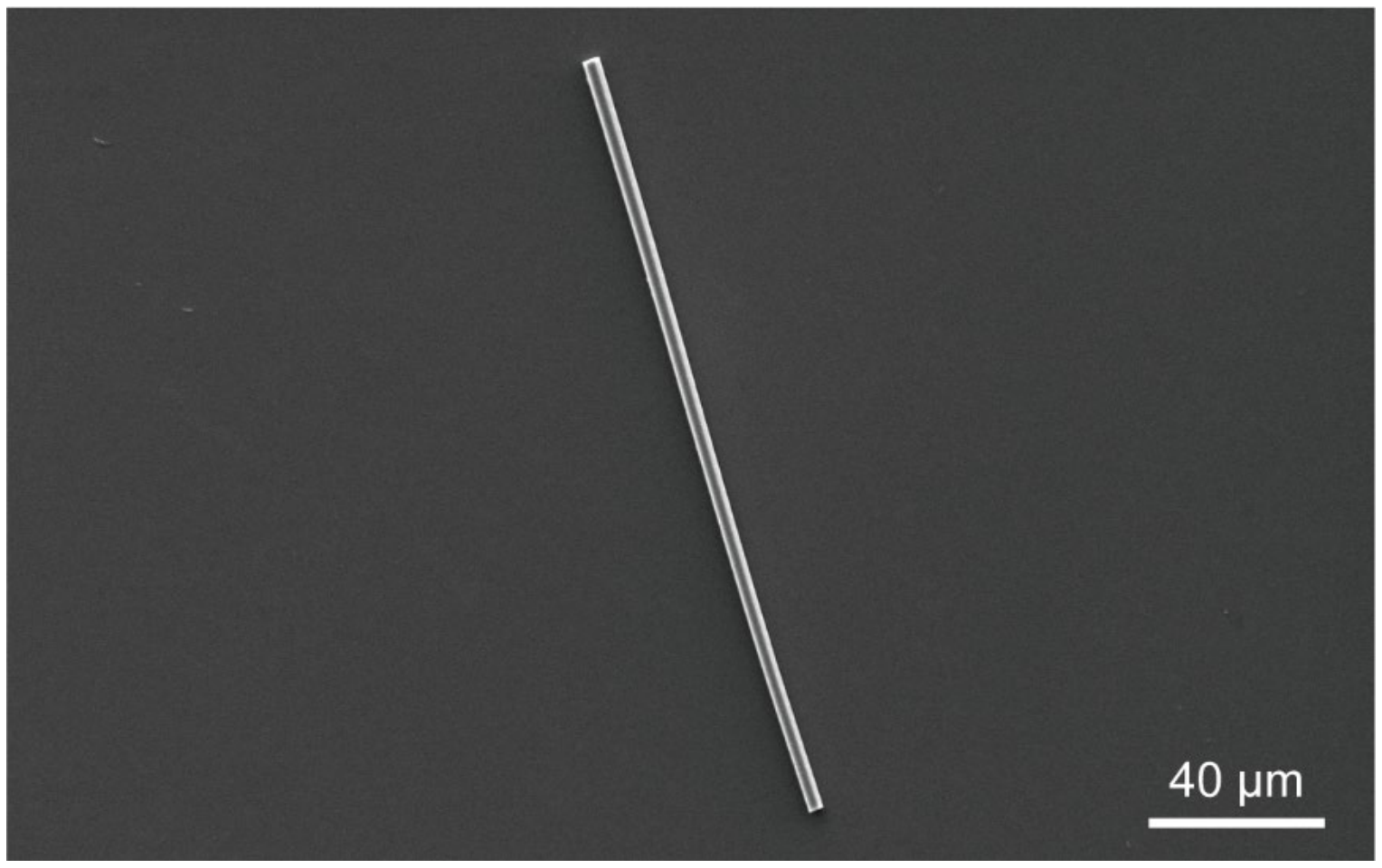

**Fig. S3.**
SEM image of $Nd_3HHTP_2$.

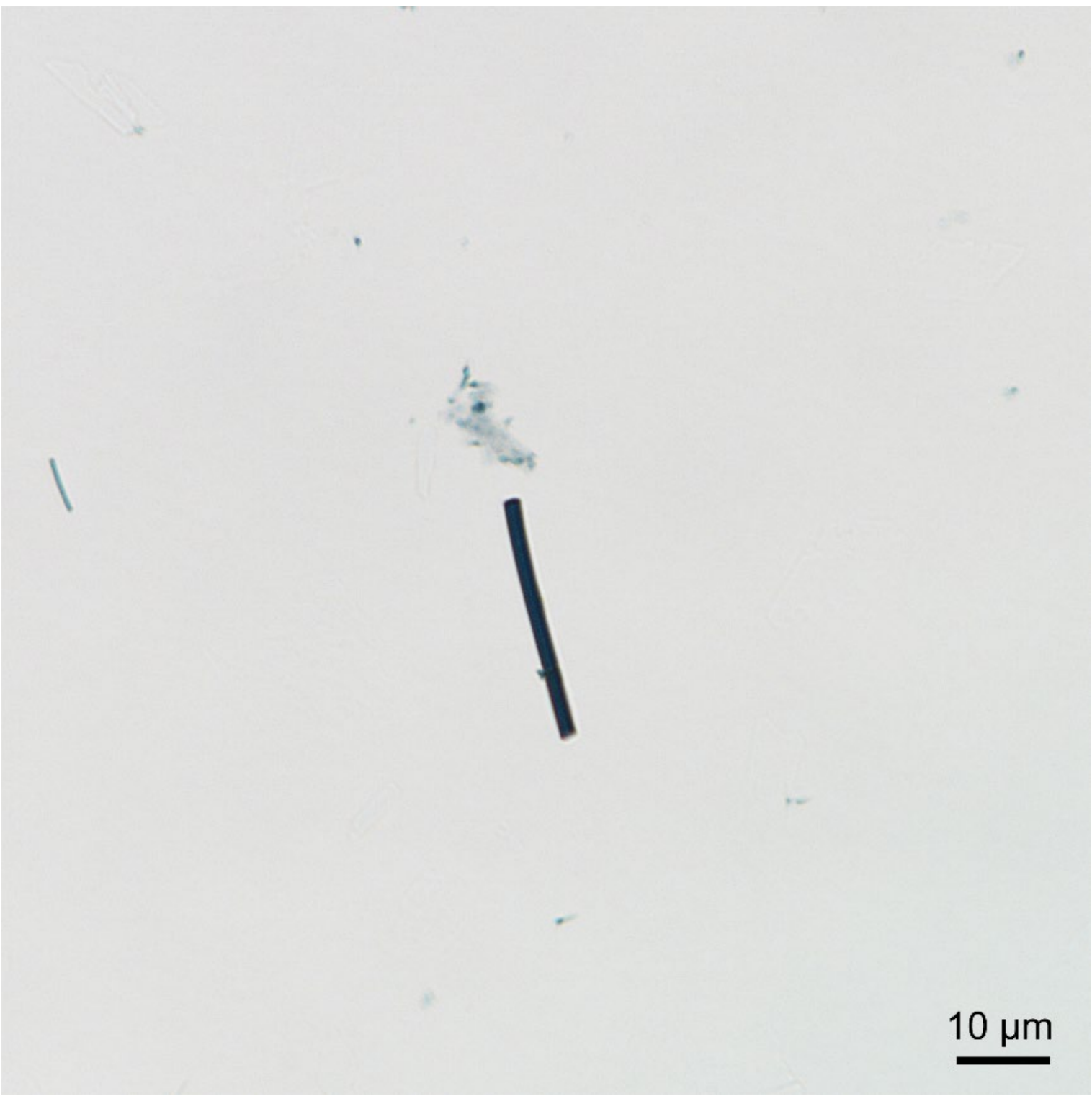

**Fig. S4.**
OM image of $Cu_3HHTP_2$.

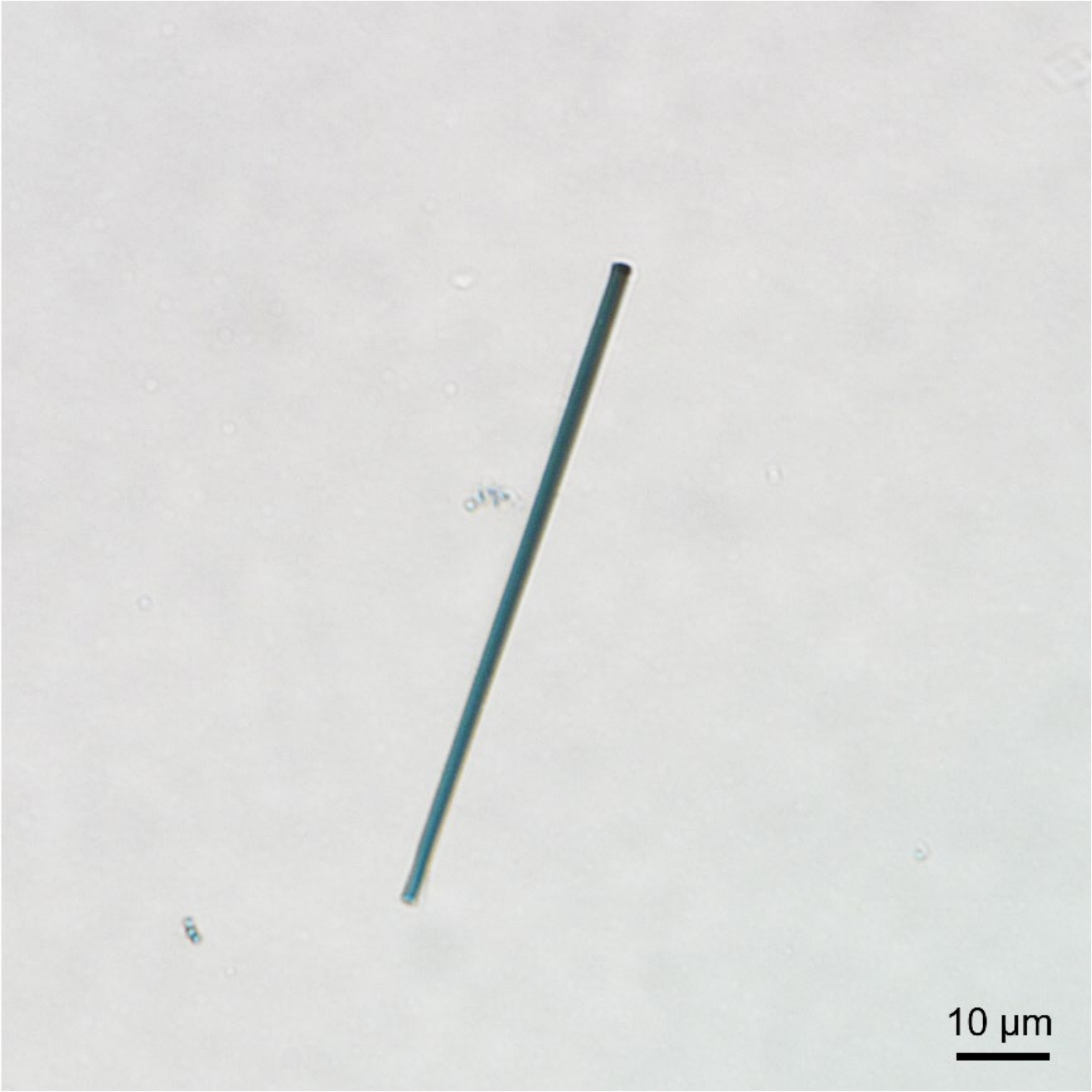

**Fig. S5.**
OM image of $Co_9HHTP_4$.

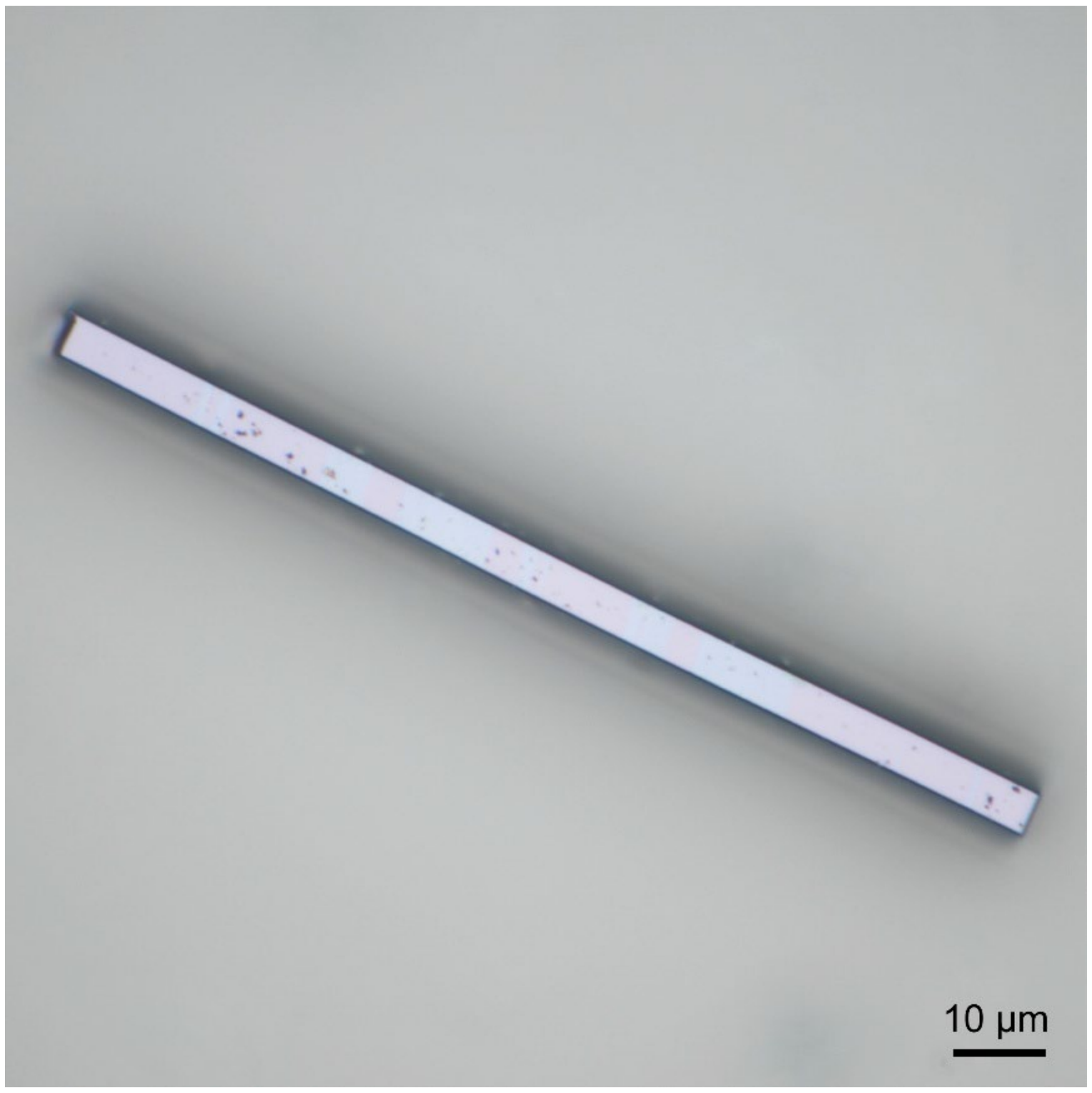

**Fig. S6.**
OM image of $Nd_3HHTP_2$.

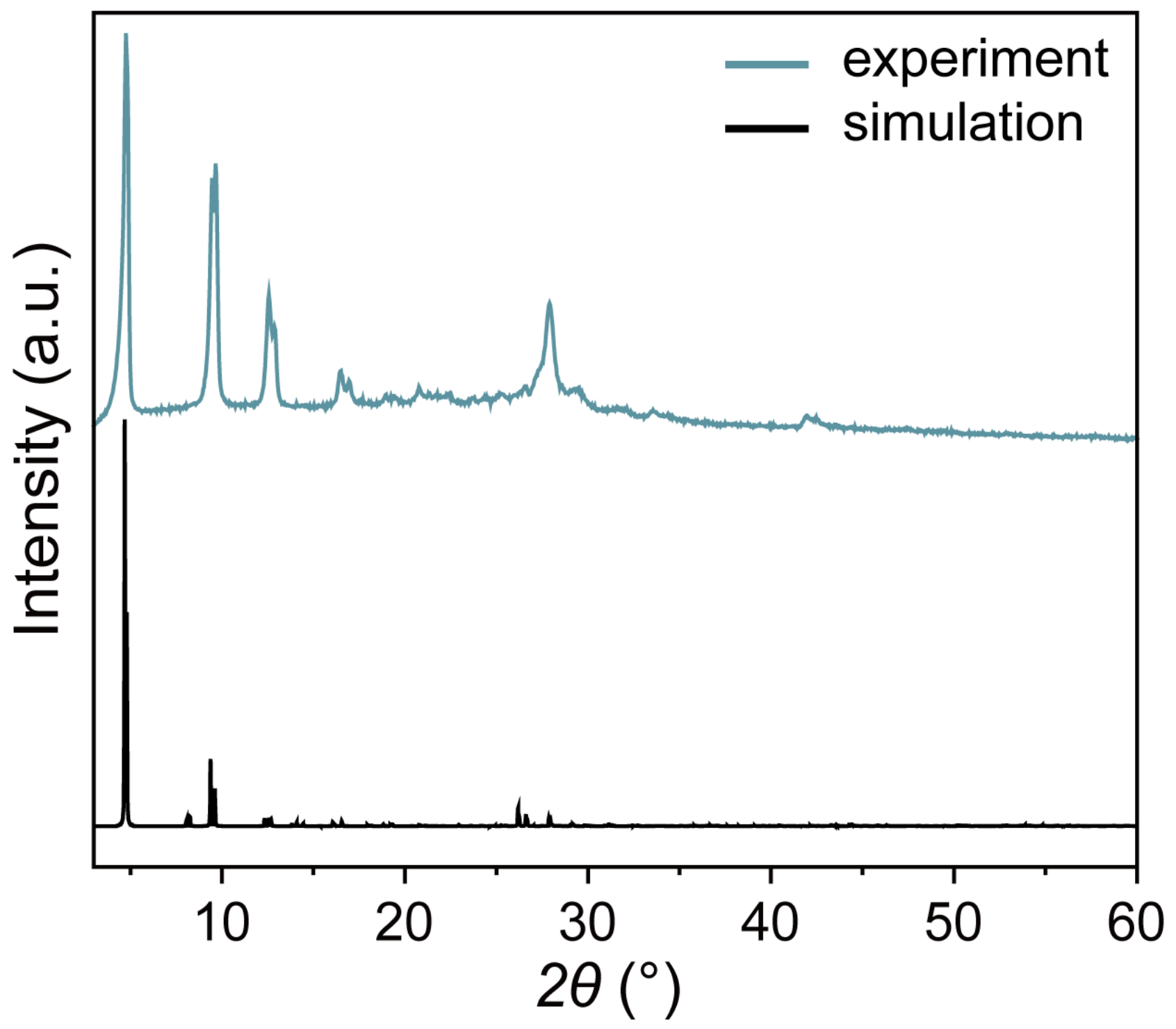

**Fig. S7.**
PXRD pattern of $Cu_3HHTP_2$, the blue solid line represents experimental data, while the black solid line represents simulated data.

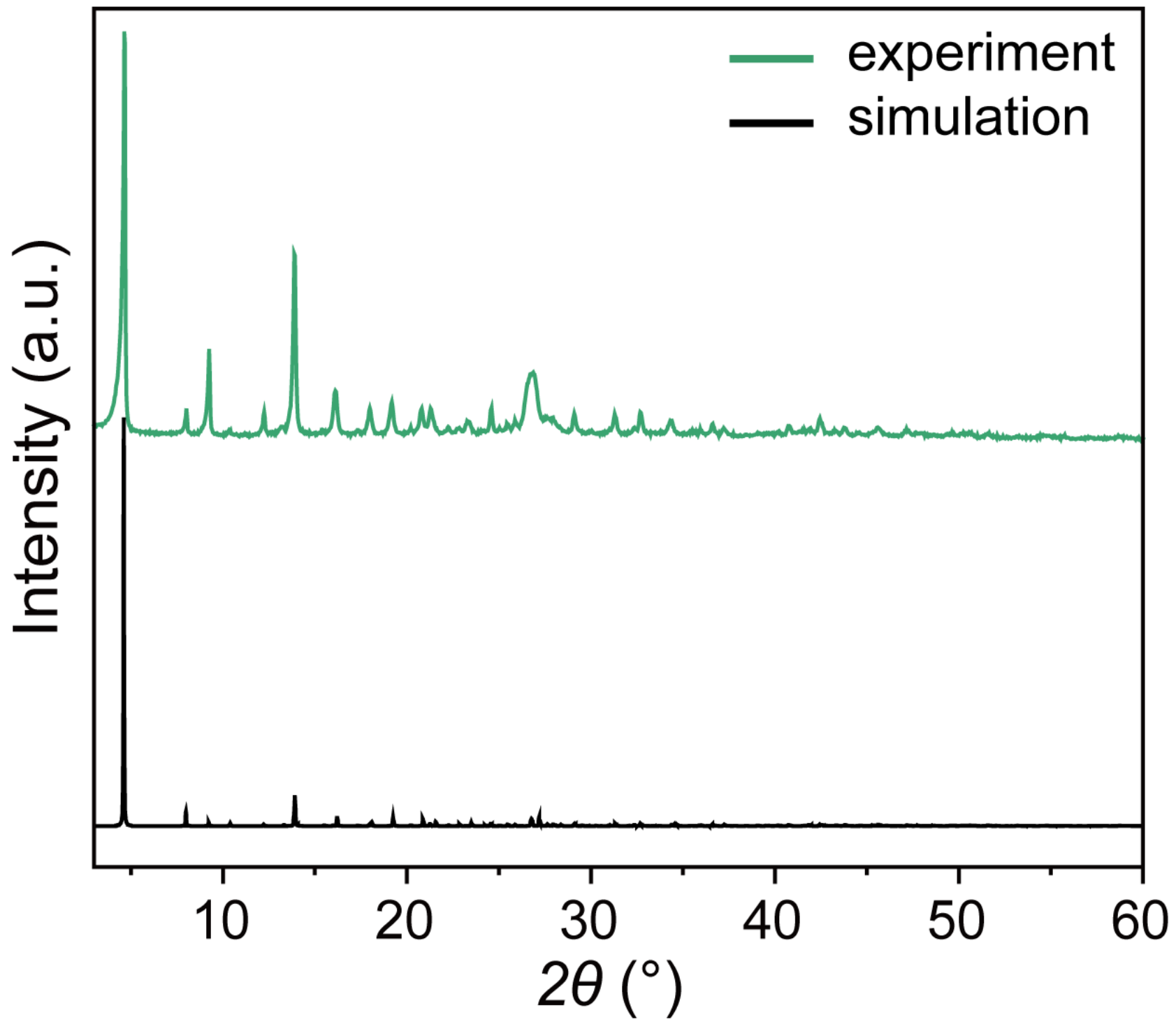

**Fig. S8.**
Experimental PXRD pattern of $Co_9HHTP_4$, the green solid line represents experimental data, while the black solid line represents simulated data.

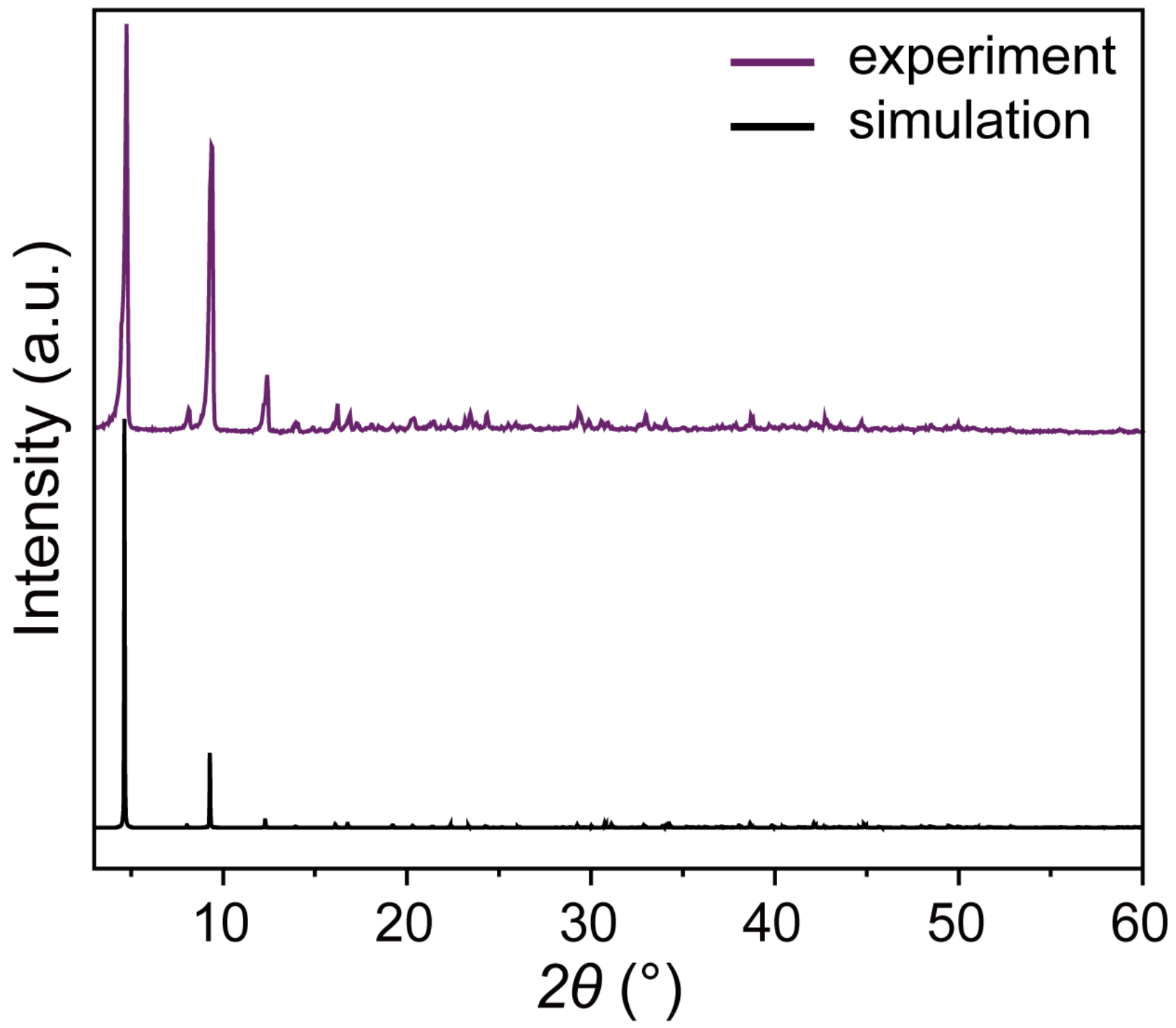

**Fig. S9.**
Experimental PXRD pattern of $Nd_3HHTP_2$, the purple solid line represents experimental data, while the black solid line represents simulated data.

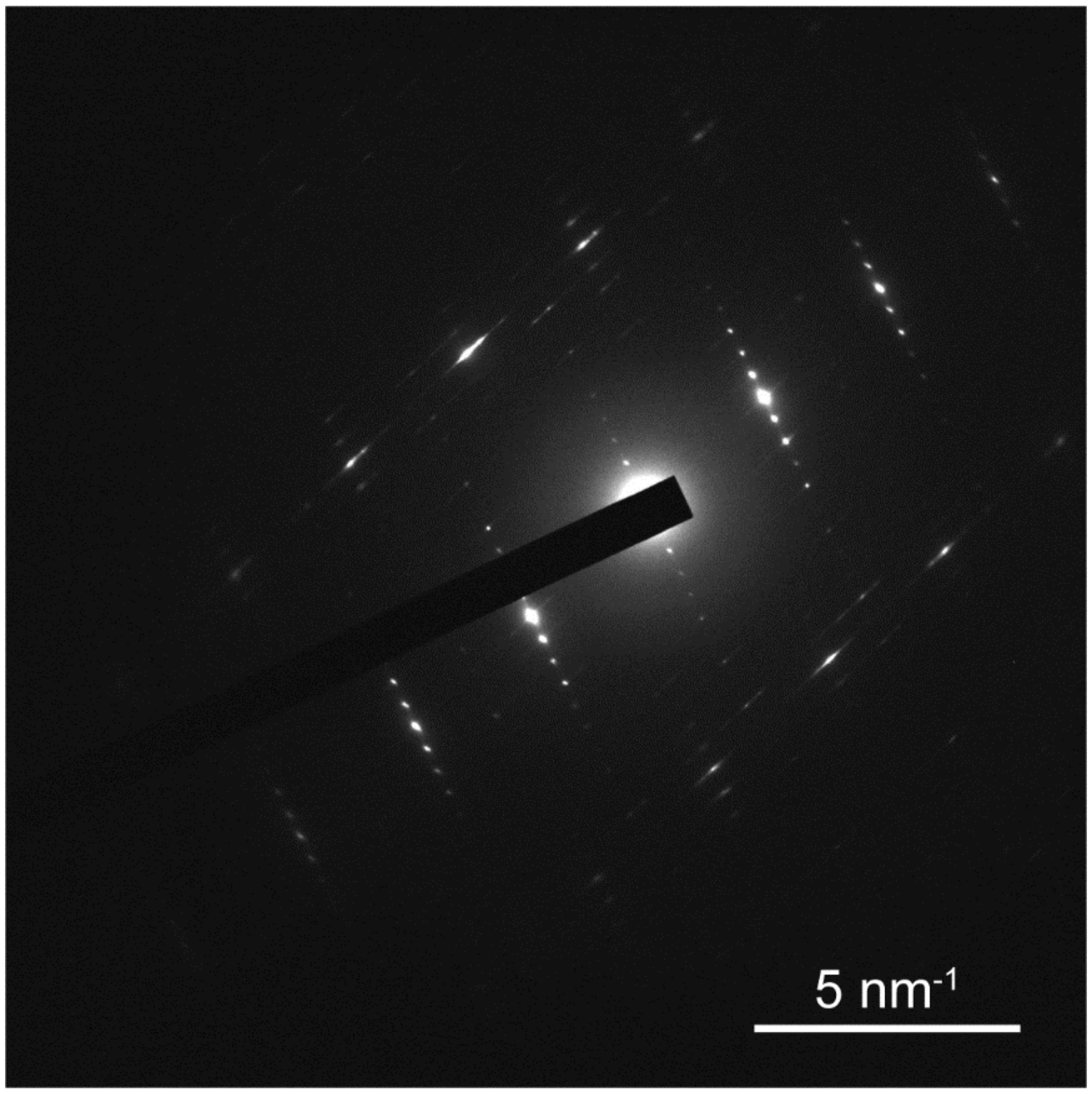

**Fig. S10.**
SAED pattern of $Cu_3HHTP_2$.

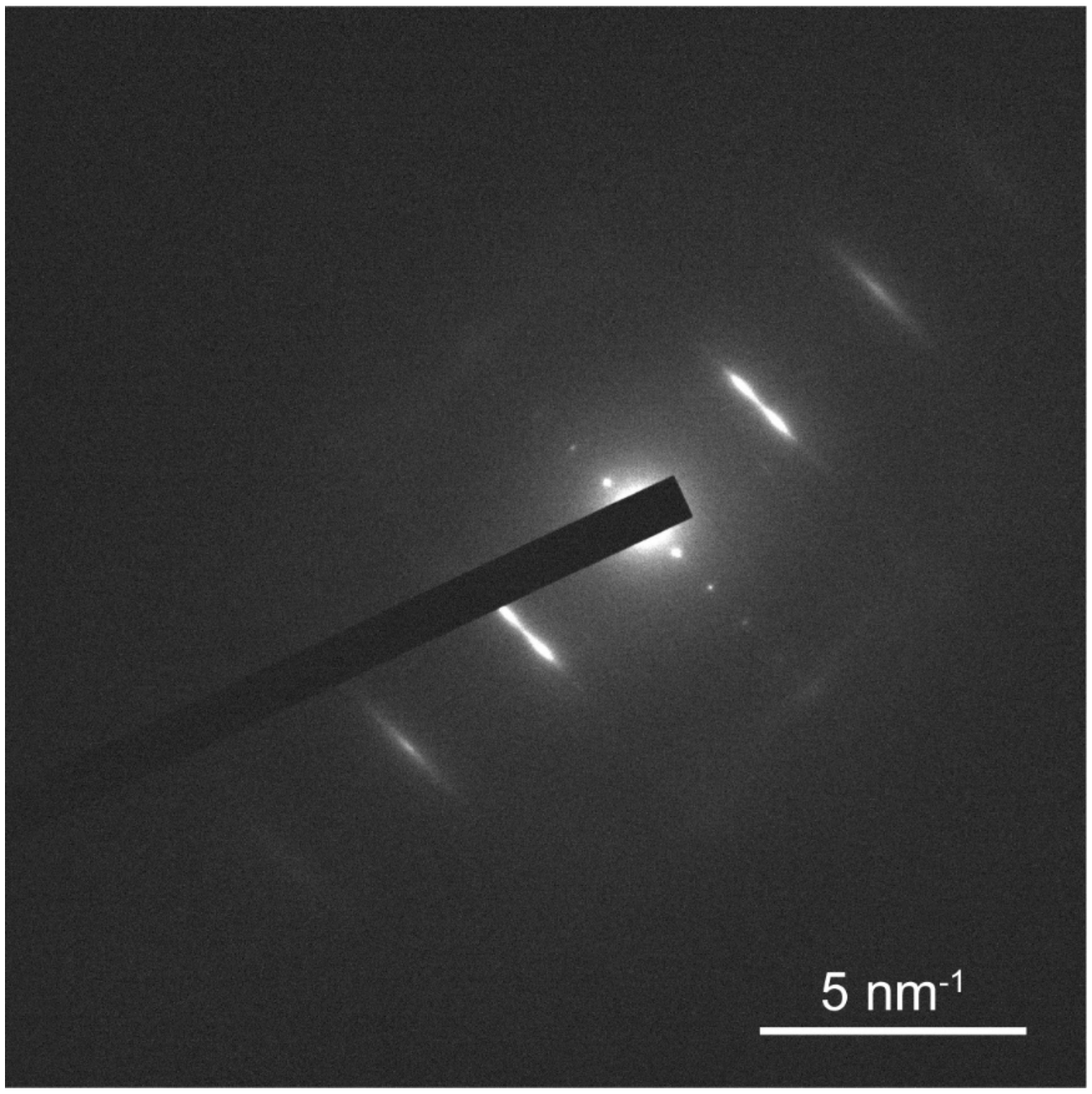

**Fig. S11.**
SAED pattern of $Co_9HHTP_4$.

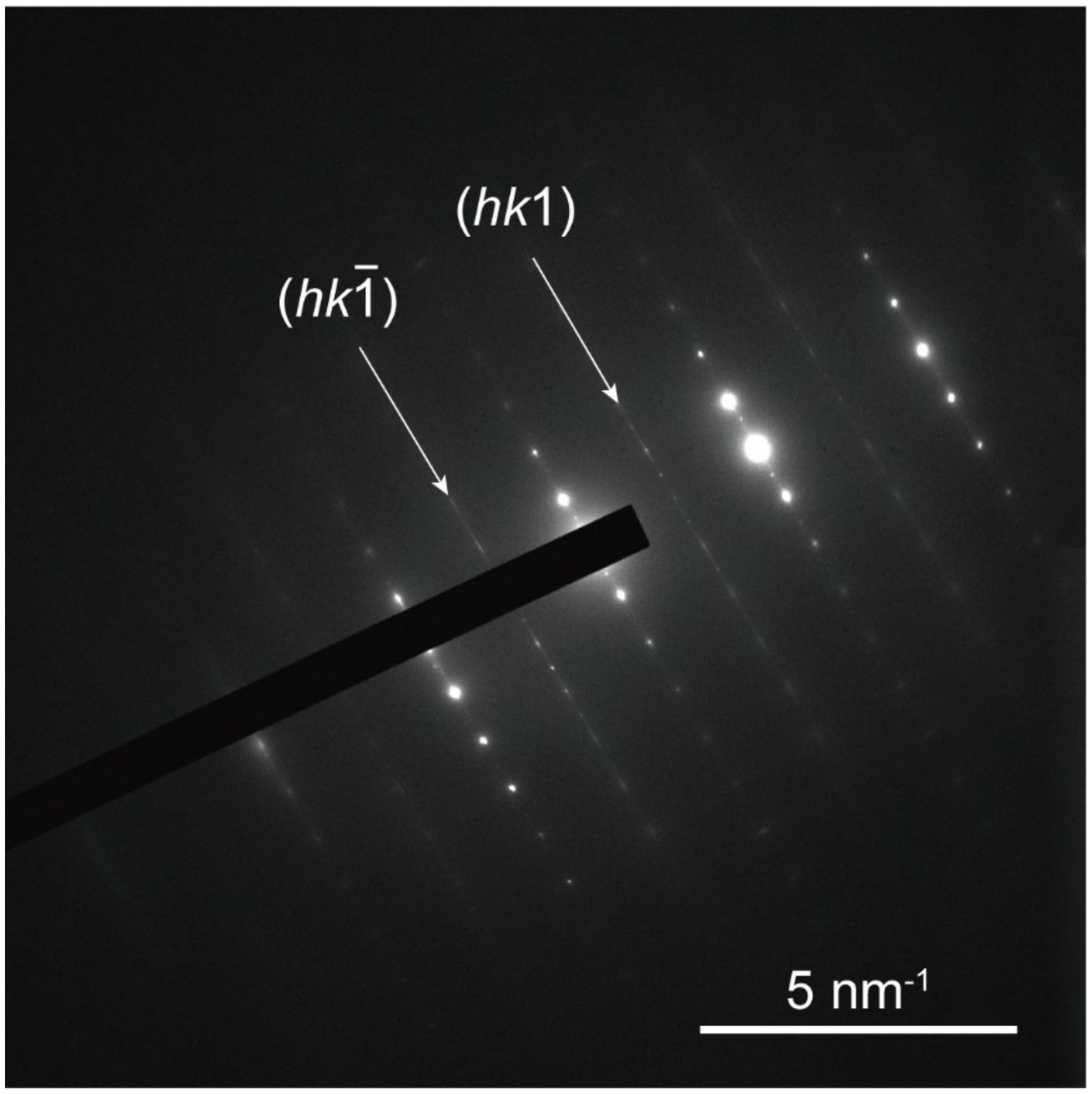

**Fig. S12.**
SAED pattern of $Nd_3HHTP_2$, when *l* is odd, the diffraction spots elongate and connect into lines, indicating the presence of diffuse reflection within the plane, which may be caused by the random occupation of $Nd^{3+}$ within the plane.

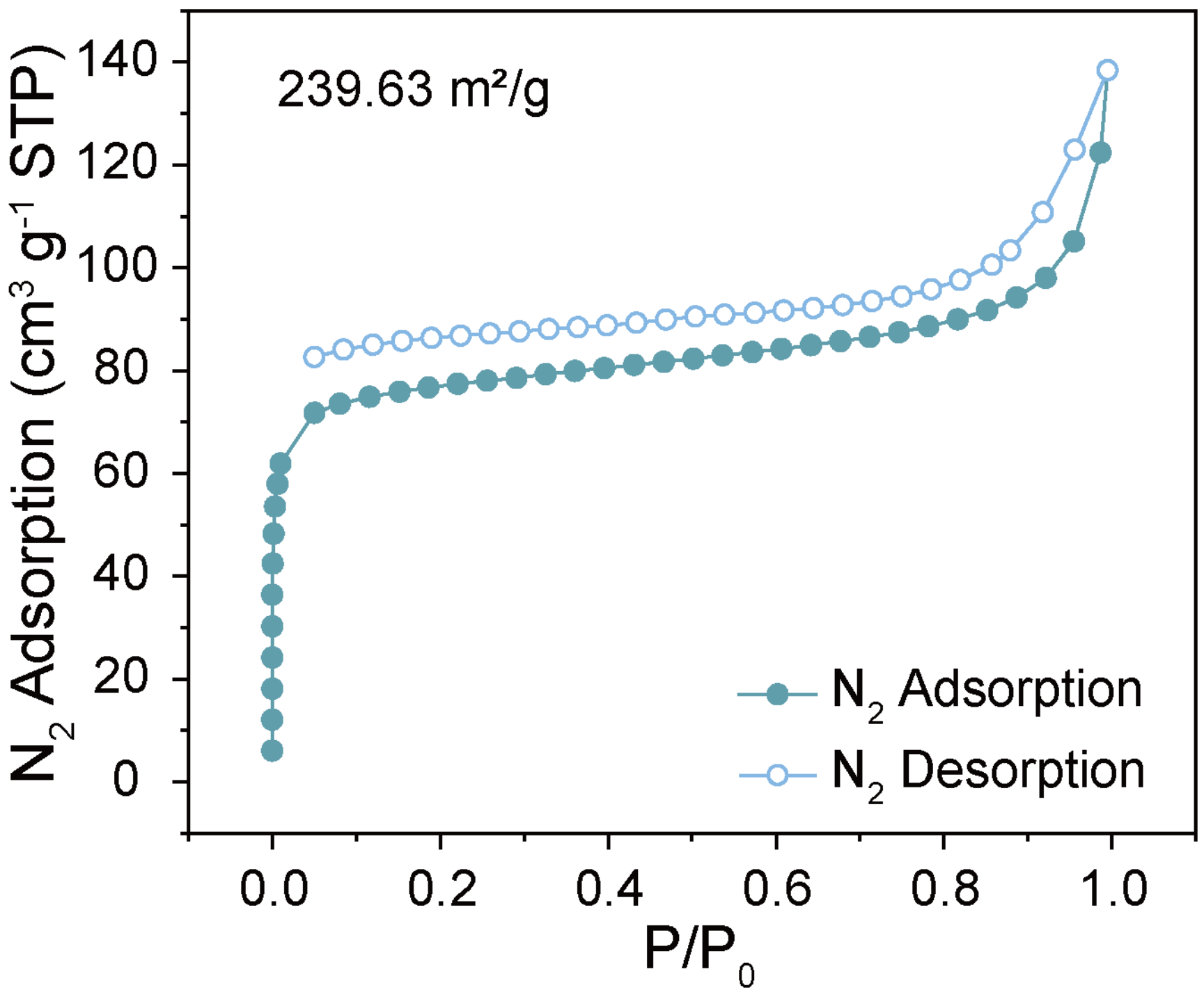

**Fig. S13.**
Nitrogen adsorption isotherms for $Cu_3HHTP_2$ at 77K. The isotherm was fit to the BET equation to give apparent BET surface areas of 239.63 m²/g.

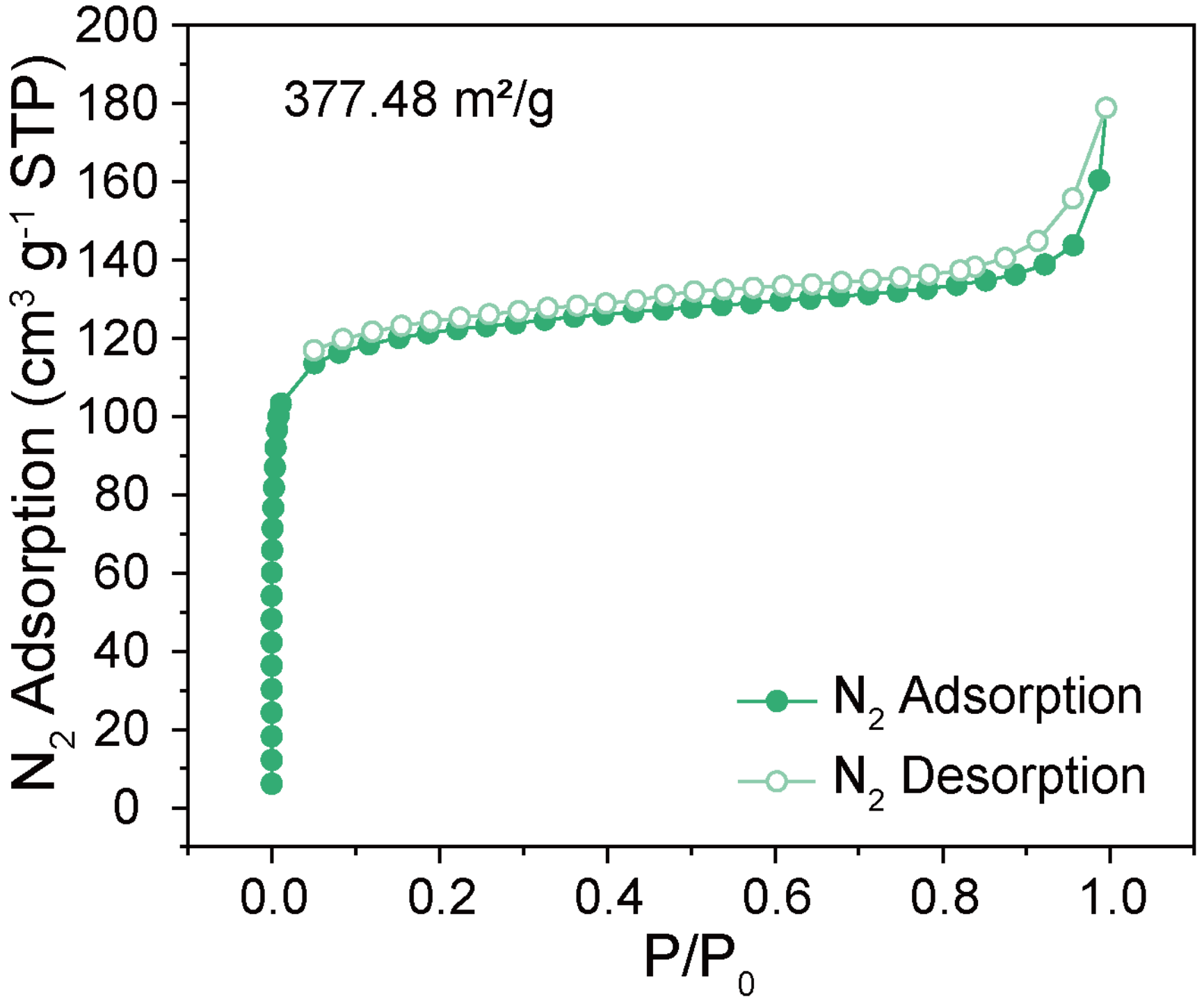

**Fig. S14.**
Nitrogen adsorption isotherms for $Co_9HHTP_4$ at 77K. The isotherm was fit to the BET equation to give apparent BET surface areas of 377.48 $m^2/g$.

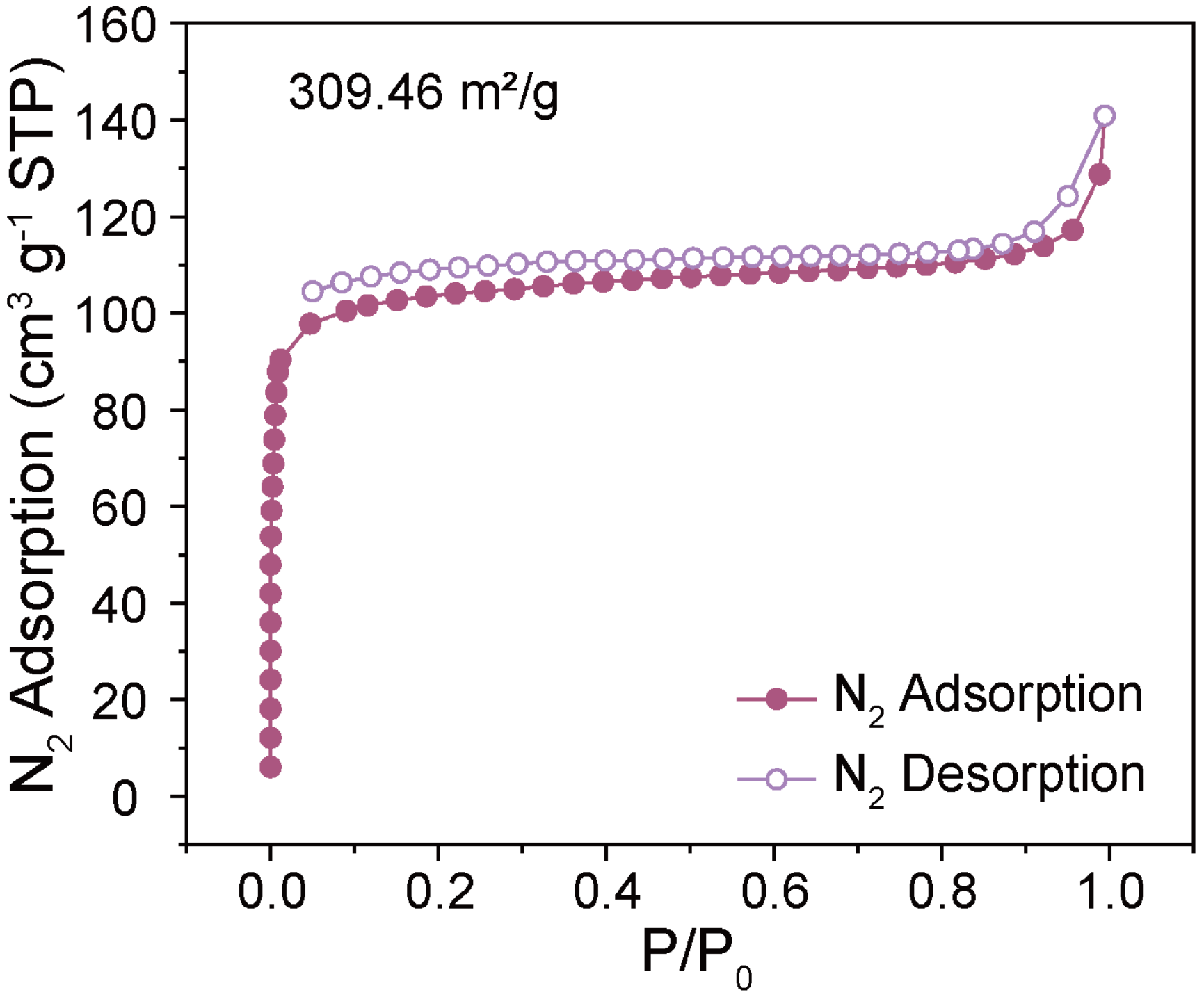

**Fig. S15.**
Nitrogen adsorption isotherms for $Nd_3HHTP_2$ at 77K. The isotherm was fit to the BET equation to give apparent BET surface areas of 309.46 m²/g.

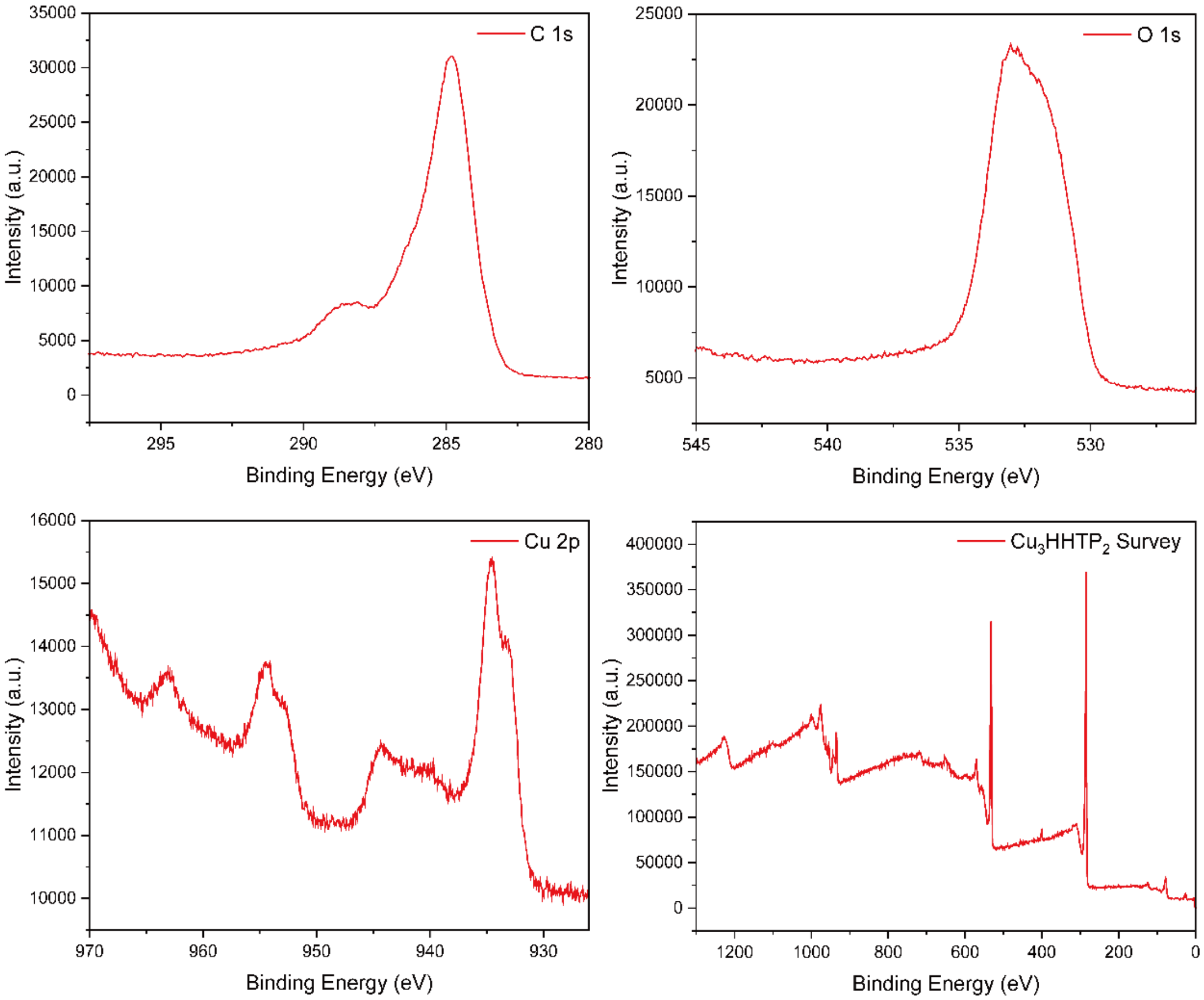

**Fig. S16.**
XPS spectra of $Cu_3HHTP_2$ survey and the corresponding C(1s), O(1s) and Cu(2p) region.

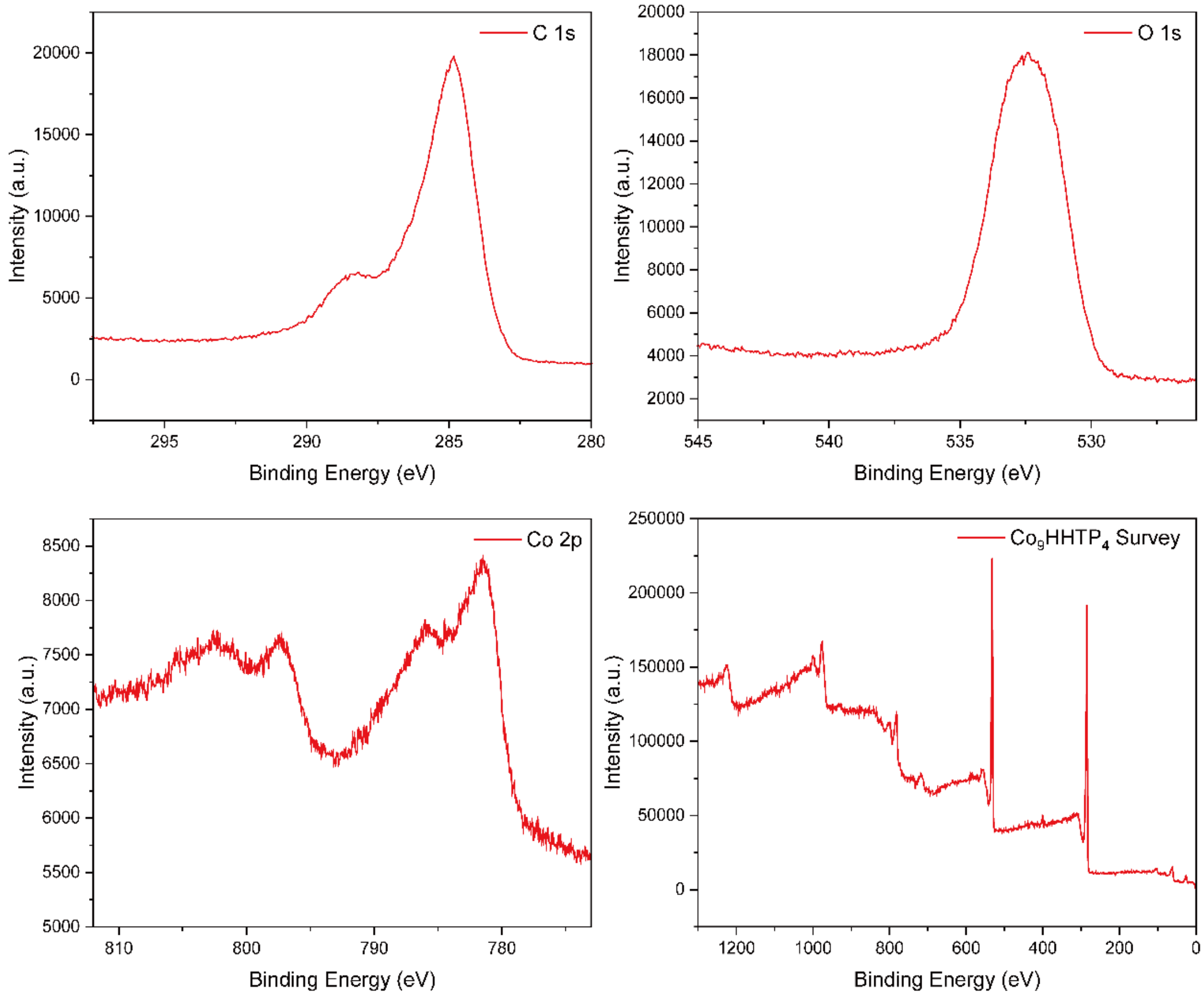

**Fig. S17.**
XPS spectra of $Co_9HHTP_4$ survey and the corresponding C(1s), O(1s) and Co(2p) region.

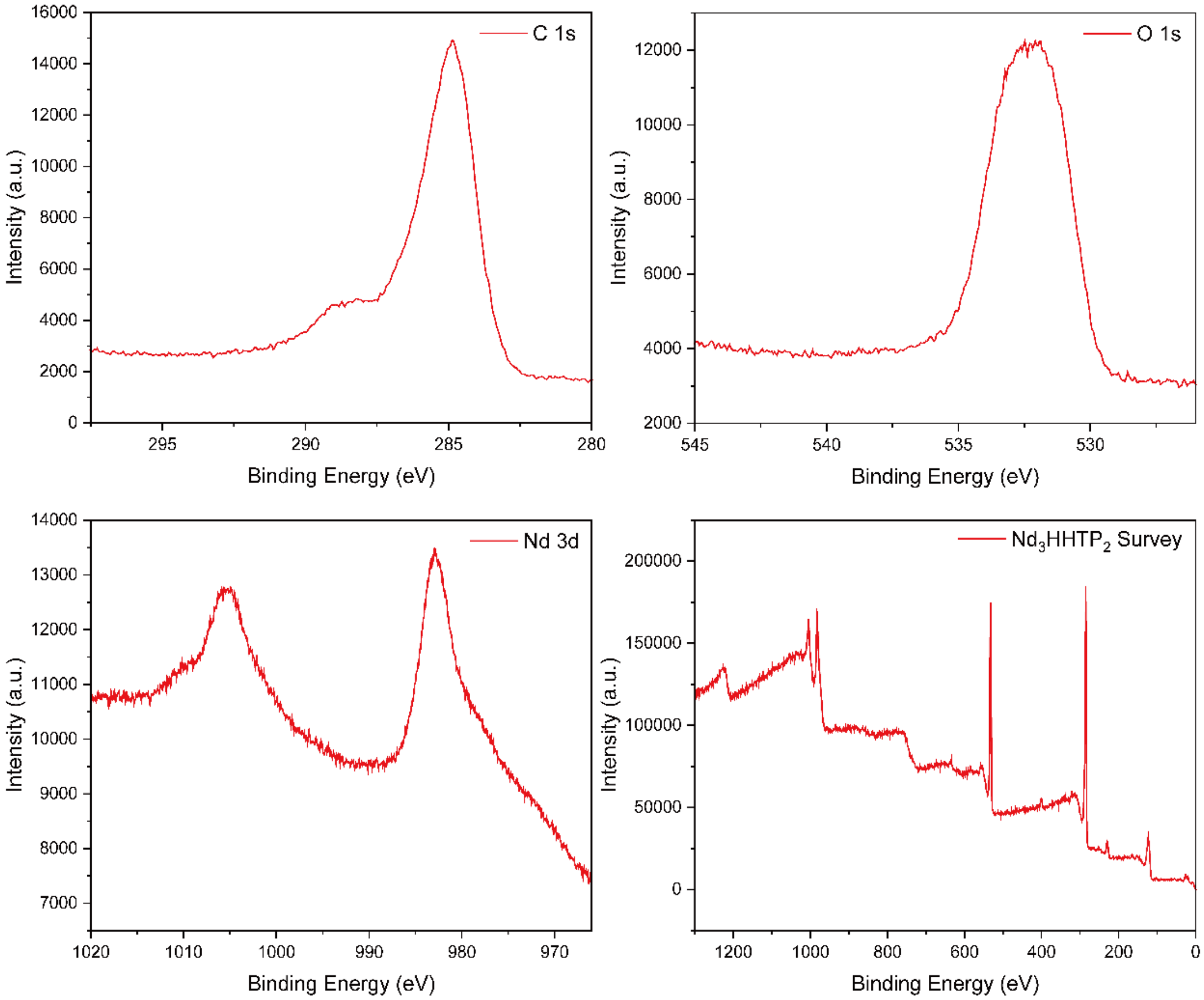

**Fig. S18.**
XPS spectra of $Nd_3HHTP_2$ survey and the corresponding C(1s), O(1s) and Nd(3d) region.

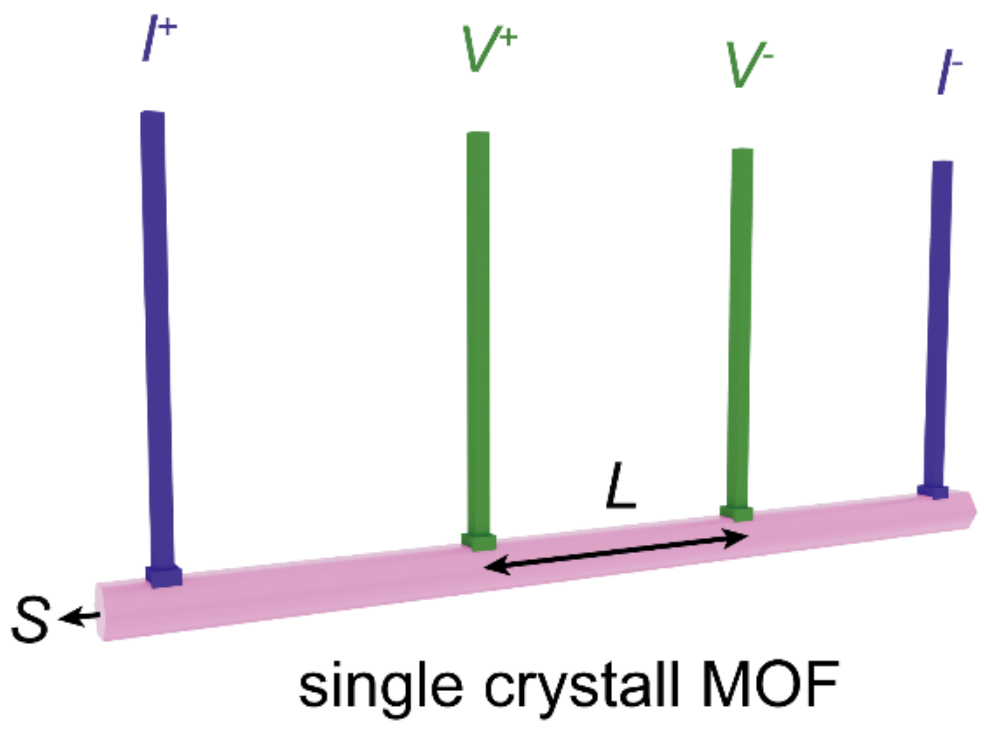

**Fig. S19.**
Schematic diagram of measuring the electrical conductivity of single crystal MOF using the four-probe method. We can use the following formula to calculate the electrical conductivity ($\sigma$) of the sample.

$$\sigma = \frac{IL}{VS}$$

In this formula, $I$ is the input current, $V$ is the voltage between the two inner contacts, $L$ is the distance between the two inner contacts, $S$ is the cross-section area through which current is passed.

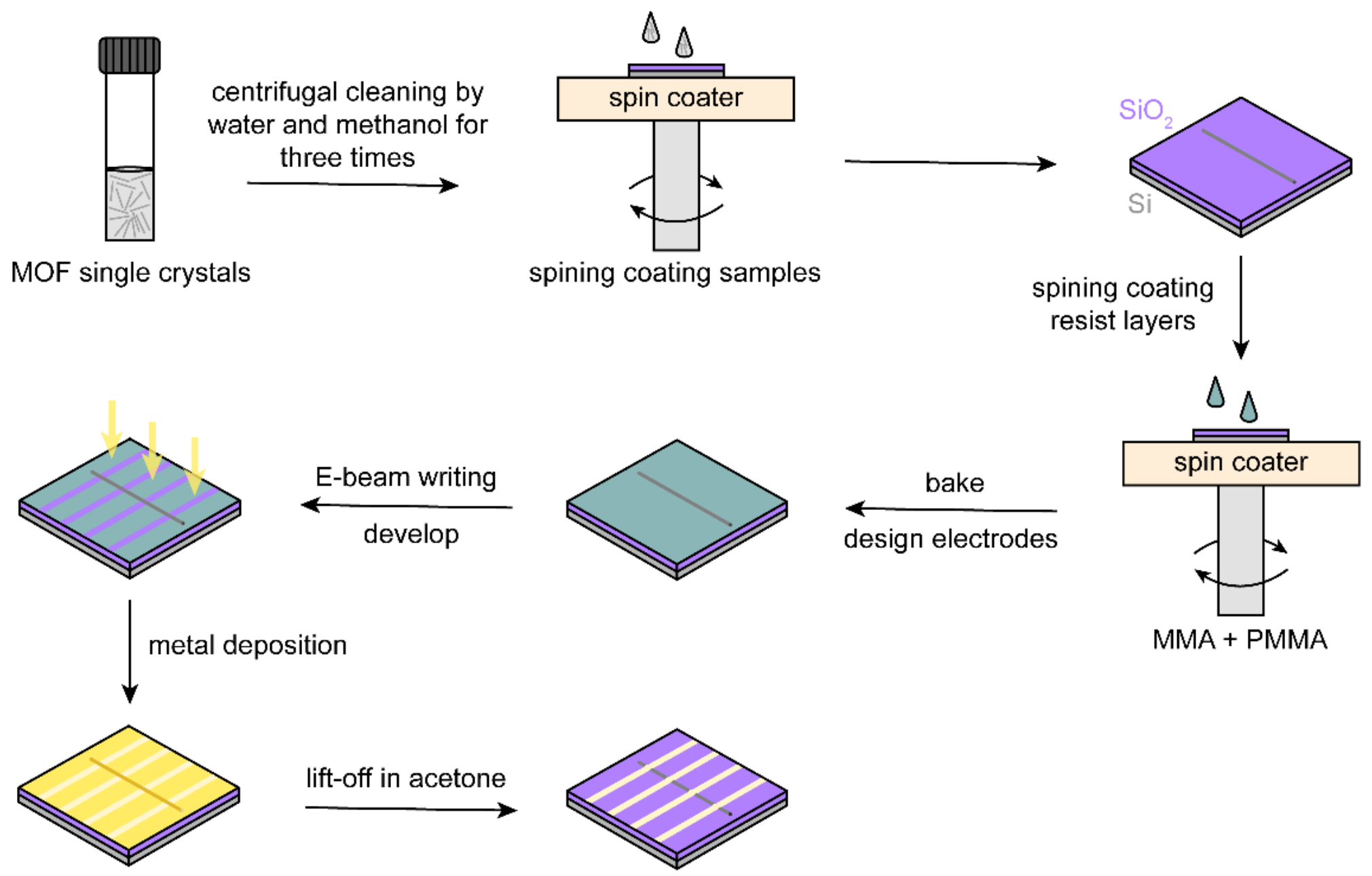

**Fig. S20.**
Diagram of EBL device fabrication process for electricity conductivity measurement of $Cu_3HHTP_2$ and $Co_9HHTP_4$.

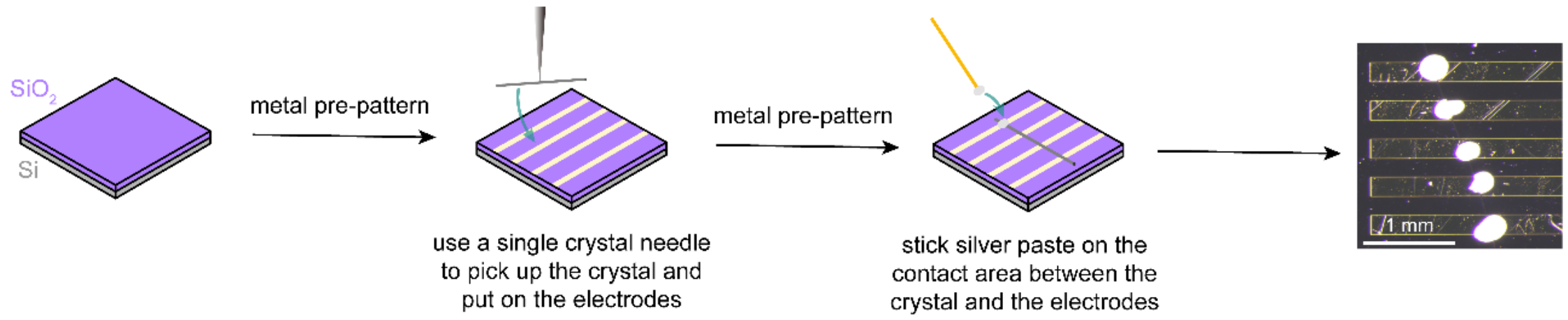

**Fig. S21.**
Diagram of device fabrication process for electricity conductivity measurement of $Nd_3HHTP_2$.

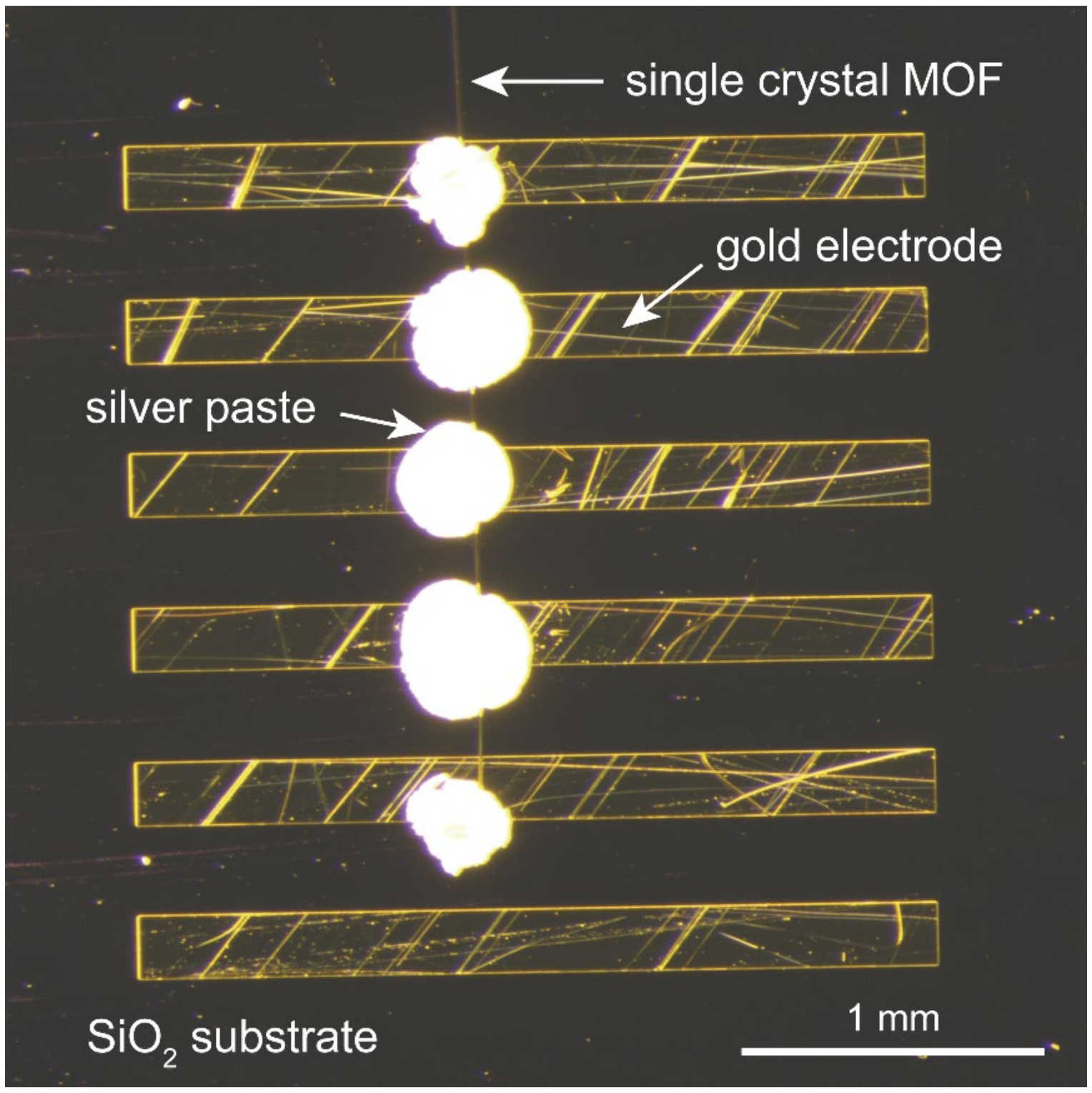

**Fig. S22.**
Electrical conductivity measurement device for $Nd_3HHTP_2$.

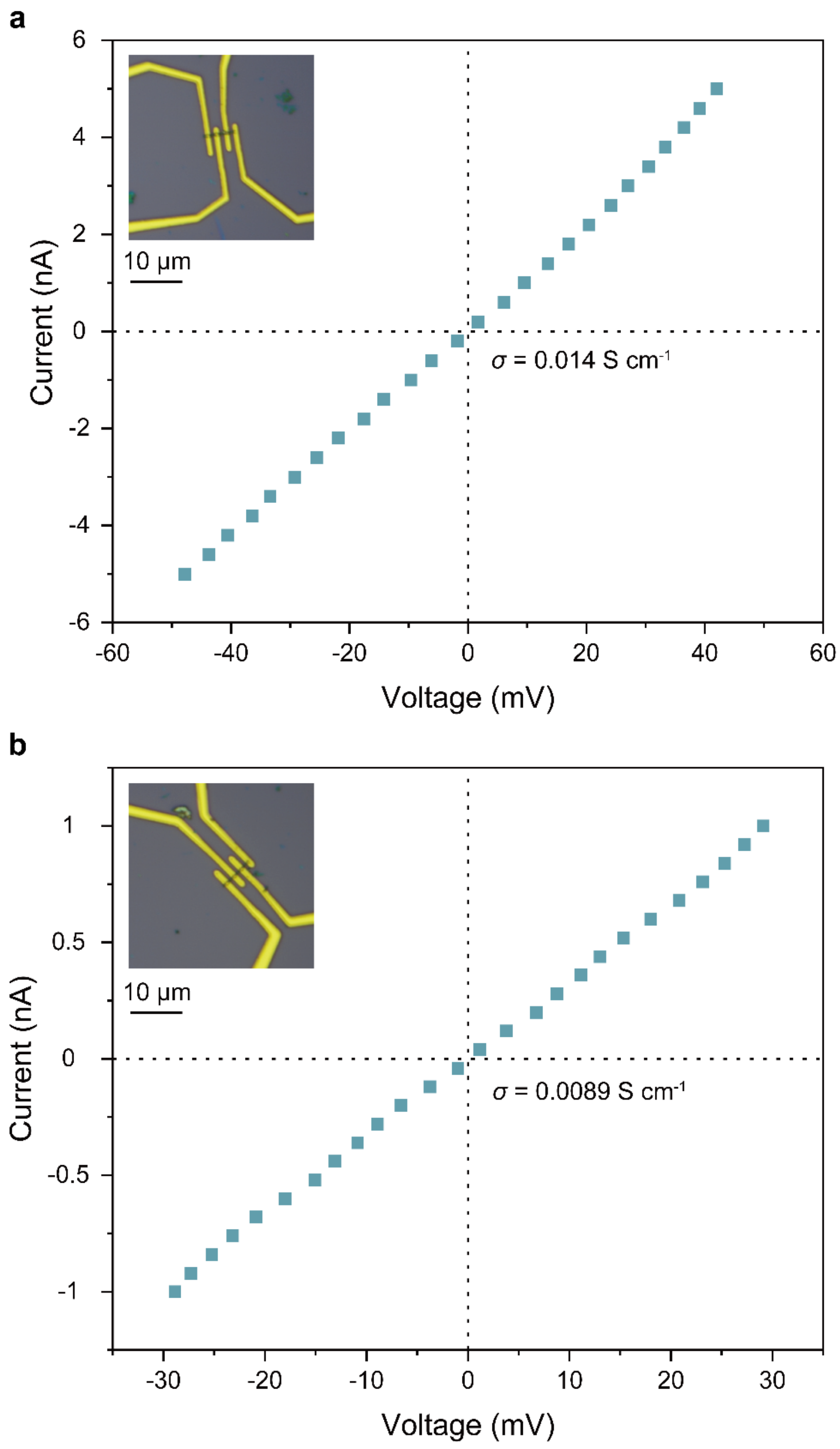

**Fig. S23.**
*I–V* curves of $Cu_3HHTP_2$.

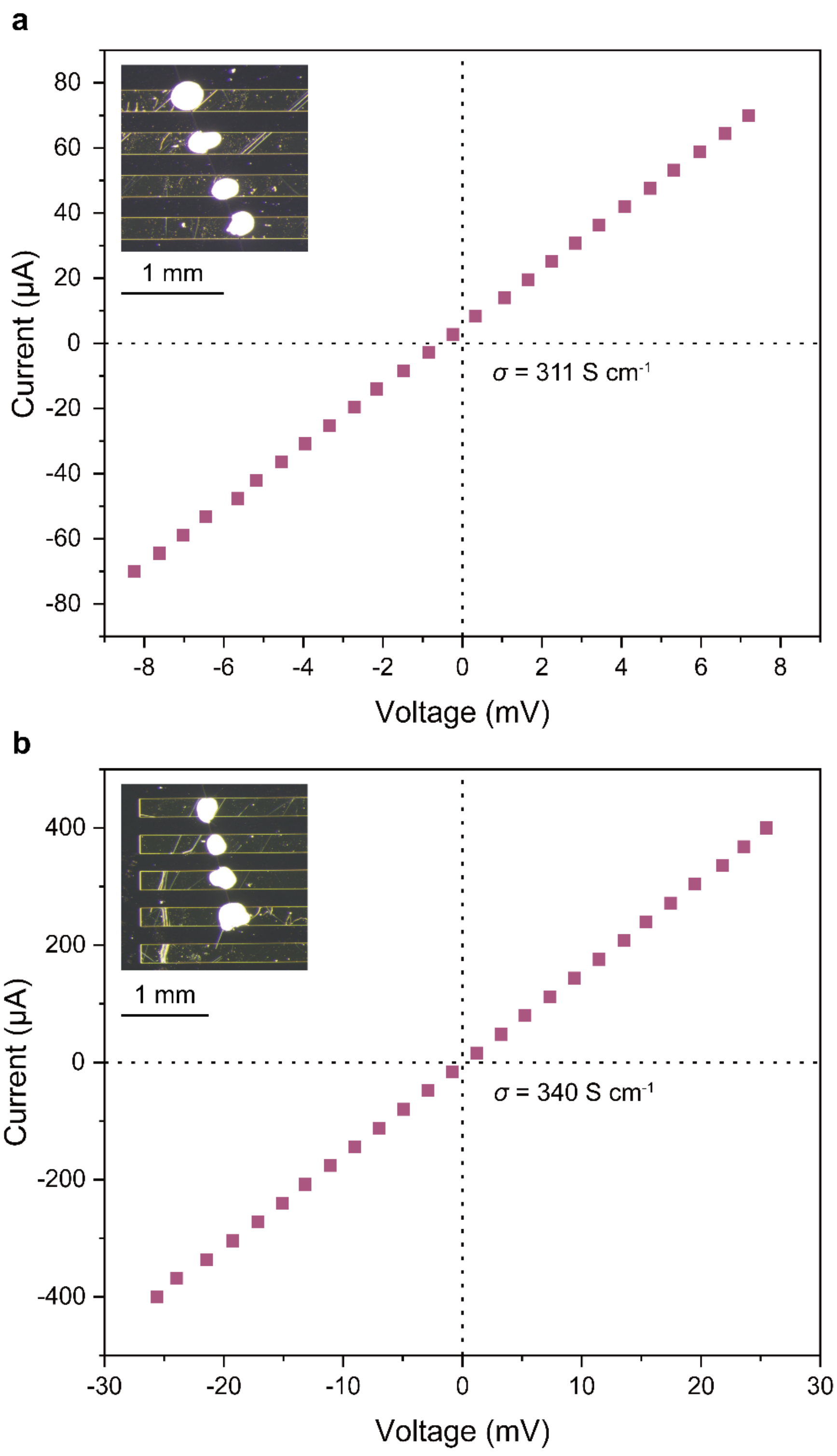

**Fig. S24.**
*I–V* curves of $Nd_3HHTP_2$.

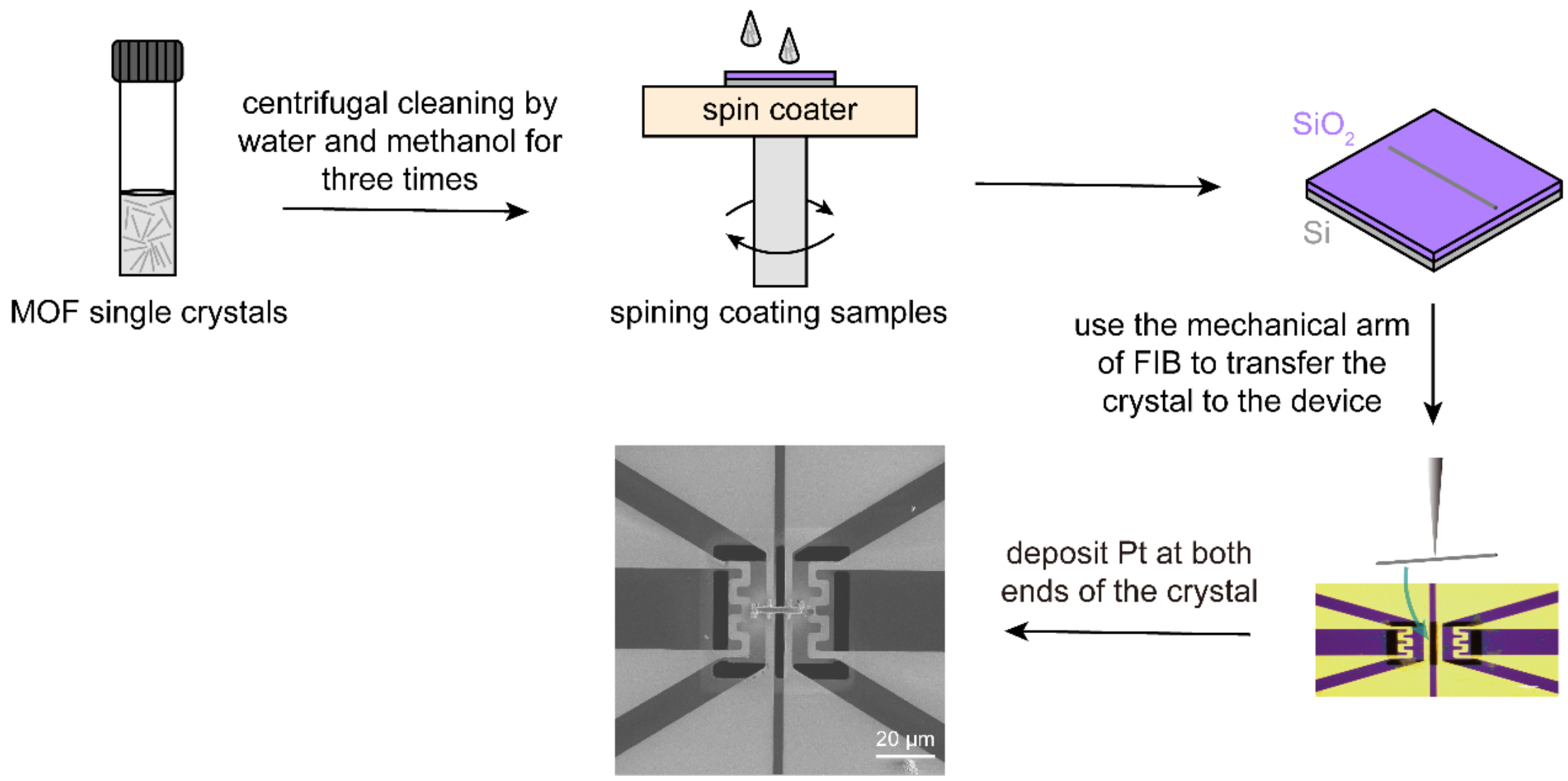

**Fig. S25.**
Diagram of device fabrication process for thermal conductivity measurement.

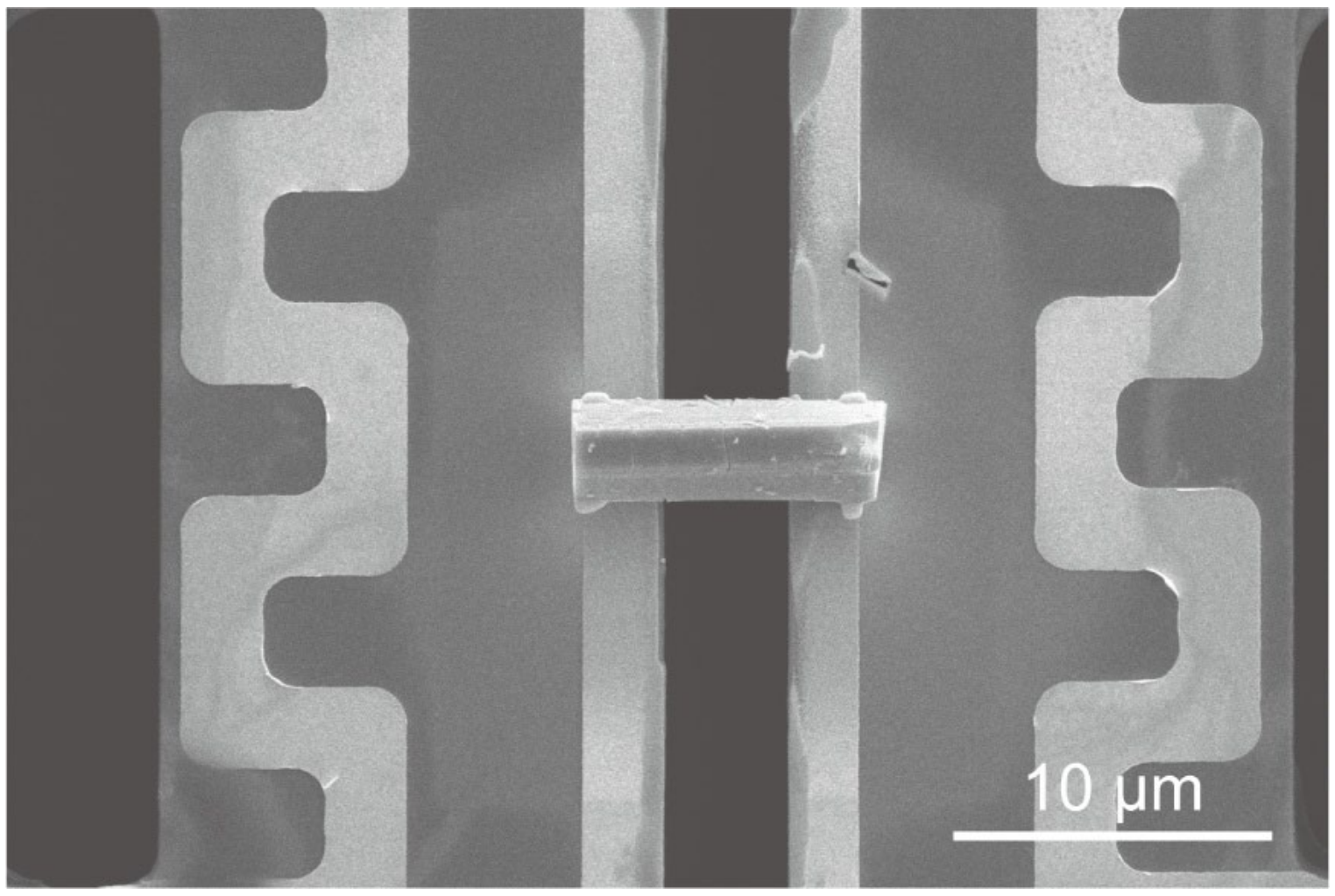

**Fig. S26.**
Thermal conductivity measurement device of $Cu_3HHTP_2$.

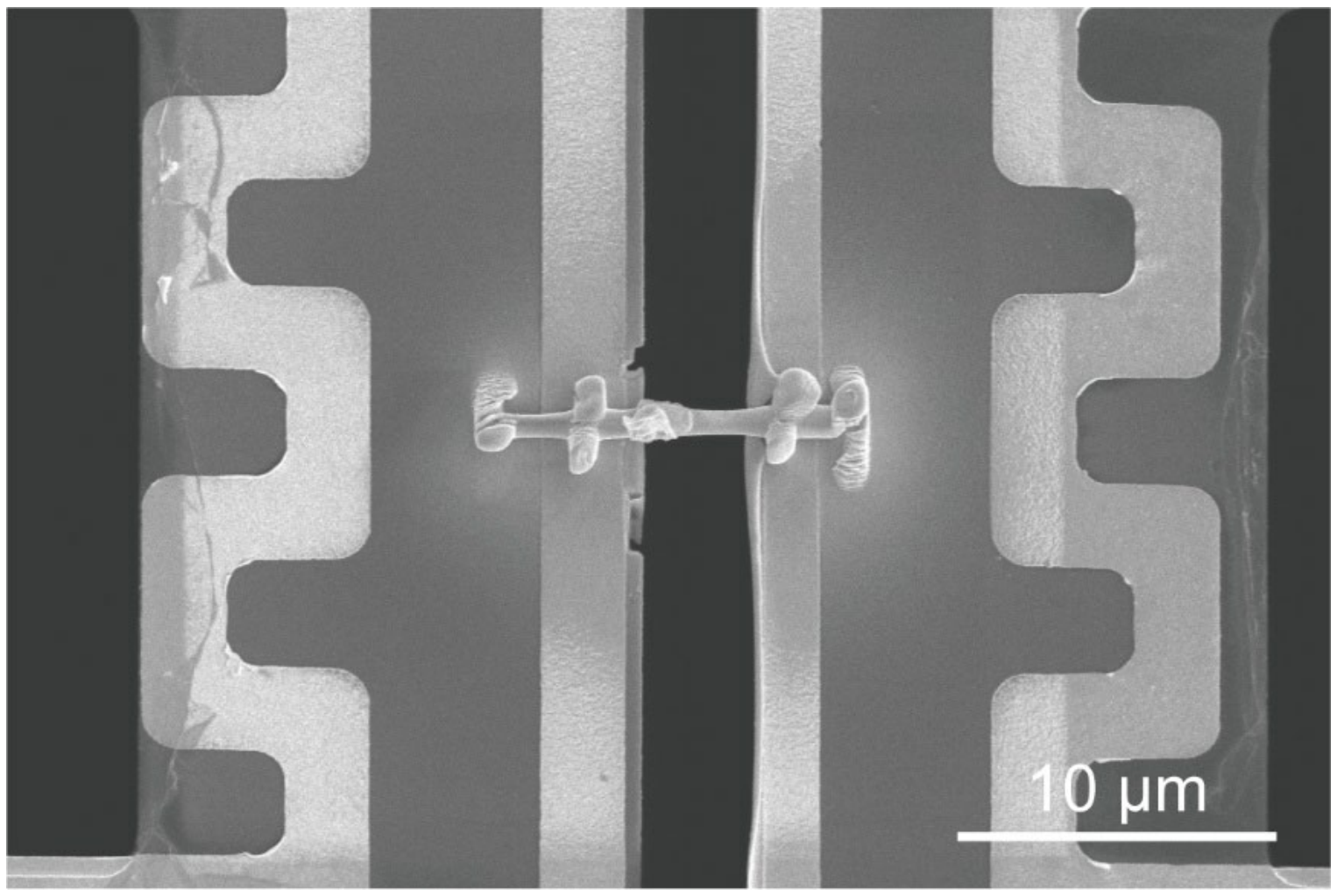

**Fig. S27.**
Thermal conductivity measurement device of $Co_9HHTP_4$.

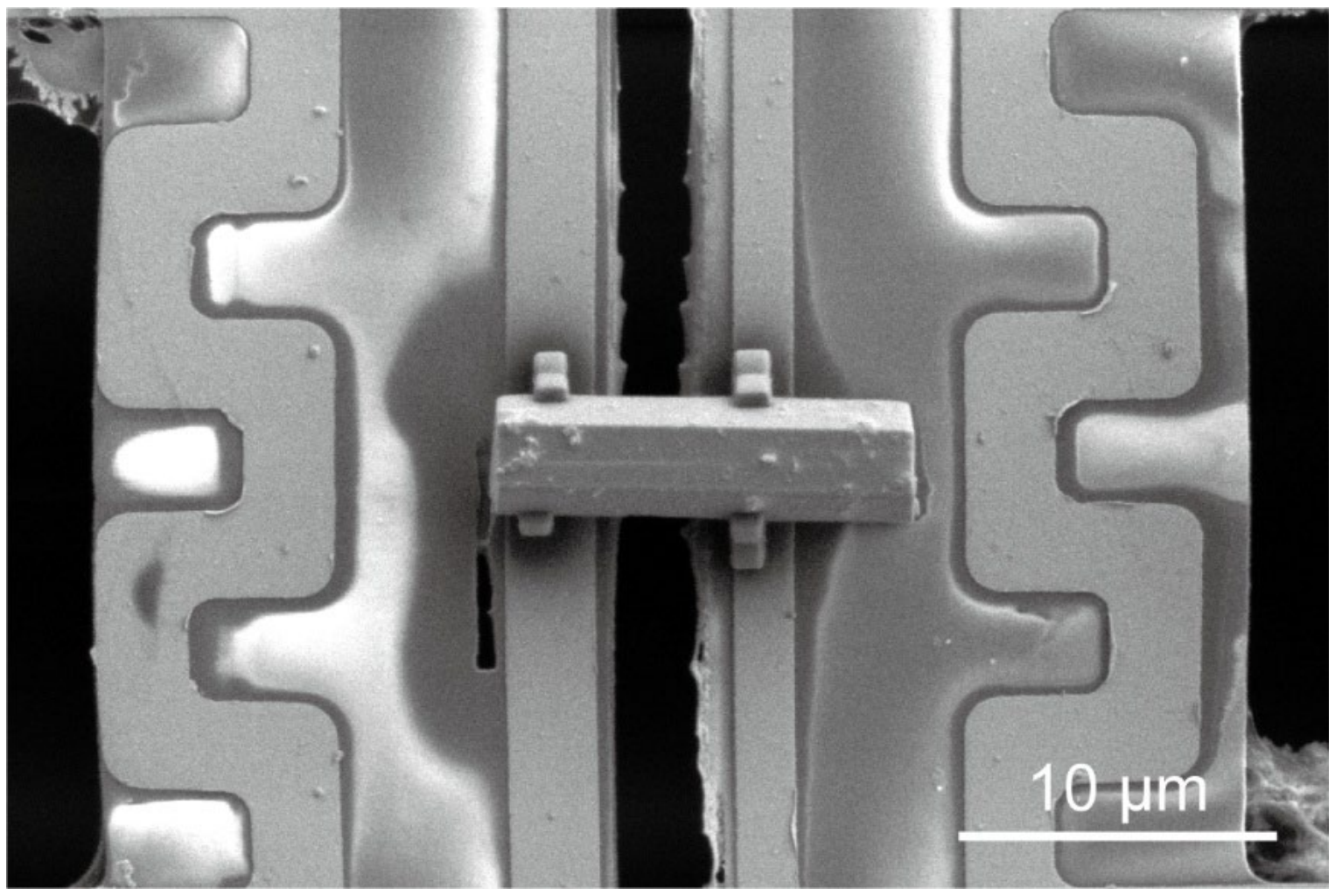

**Fig. S28.**
Thermal conductivity measurement device of $Nd_3HHTP_2$.

**Table. S1.**
The electrical conductivity and thermal conductivity of the materials involved in Fig. 4f, the bold ones are the materials measured in this work.

| Materials | Electrical conductivity ($S\ cm^{-1}$) | Thermal conductivity ($W\ m^{-1}\ K^{-1}$) |
|---|---|---|
| $Cu_2Se$ [6,7] | 100 | 0.37 |
| SnSe [8] | 88.8 | 0.35 |
| $Sb_2Te_3$ [9] | 782.5 | 0.91 |
| CNTs [10] | 13.6 | 0.28 |
| CuI [11] | 143 | 0.55 |
| $Ag_2Se$ [12] | 497 | 0.478 |
| HgSe [13] | 20.5 | 0.24 |
| HgTe [13] | 0.35 | 0.17 |
| PEDOT:PSS [14] | 639 | 0.23 |
| PEDOT:Tos [15] | 80 | 0.33 |
| EtOPV-co-PV [16] | 2.9 | 0.66 |
| Ppy:Tos [17] | 170 | 0.2 |
| TCNQ-$Cu_3(BTC)_2$ [18] | 0.0045 | 0.27 |
| $Ni_3(HITP)_2$ [19] | 58.8 | 0.21 |
| $Cu_{4/3}$-BHT [20] | 270 | 0.72 |
| $Cu_{5.5/3}$-BHT [20] | 220 | 0.59 |
| **$Cu_3HHTP_2$** | **0.7** | **0.075** |
| **$Co_9HHTP_4$** | **0.69** | **0.194** |
| **$Nd_3HHTP_2$** | **398** | **0.148** |

**Table. S2.**

Crystal data of $Cu_3HHTP_2$ single crystals.

| | |
|---|---|
| Empirical formula | $Cu_3C_{36}H_{12}O_{12}$ |
| Formula weight | 827.12 |
| Crystal system | Orthorhombic |
| Space group | *Cmcm* |
| Unit cell dimensions | $a$ = 36.8075 Å $\alpha$ = 90° |
| | $b$ = 21.9562 Å $\beta$ = 90° |
| | $c$ = 6.8031 Å $\gamma$ = 90° |
| Volume | 5497.9 $Å^3$ |
| Z | 4 |

The crystal data of $Cu_3HHTP_2$ was obtained by Pawley fitting of experimental PXRD pattern.

**Table. S3.**

Crystal data of $Co_9HHTP_4$ single crystals.

| | |
|---|---|
| Empirical formula | $Co_9C_{72}H_{24}O_{80}$ |
| Formula weight | 2699.28 |
| Crystal system | Trigonal |
| Space group | *P-3c1* |
| Unit cell dimensions | $a$ = 22.1303 Å $\alpha$ = 90° |
| | $b$ = 22.1303 Å $\beta$ = 90° |
| | $c$ = 13.3103 Å $\gamma$ = 120° |
| Volume | 5645.1 Å$^3$ |
| Z | 2 |

The crystal data of $Co_9HHTP_4$ was obtained from the structure reported in previous work. [21]